\documentclass[10pt]{aastex}
\usepackage{emulateapj5,apjfonts}

\newcommand{\sn}	{\smallskip\noindent}%
\newcommand{\mn}	{\medskip\noindent}%
\newcommand{\eg}	{\mbox{e.g.}}%
\newcommand{\ie}	{\mbox{i.e.}}%
\newcommand{\etal}	{\mbox{et al.\ }}%
\newcommand{\aka}	{\mbox{a.k.a.\ }}%
\newcommand{\HST}	{\emph{HST}}%
\newcommand{\Astro}	{\emph{Astro}}%
\newcommand{\UIT}	{\emph{UIT}}%
\newcommand{\kmsMpc}	{\mbox{km s$^{-1}$ Mpc$^{-1}$}}%
\newcommand{\magarc}	{\ensuremath{\mbox{mag arcsec}^{-2}}}%
\newcommand{\SB}	{\ensuremath{\mu_{\mbox{\scriptsize\sl F300W}}}}%
\newcommand{\re}	{\ensuremath{r_e}}%
\newcommand{\HI}	{\mbox{H\,{\sc i}}}%
\newcommand{\HII}	{\mbox{H\,{\sc ii}}}%
\newcommand{\tsim}	{\ensuremath{\sim}}%
\newcommand{\lsim}	{\makebox[15pt]
           {\mbox{{\raisebox{-3pt}{\tsim}}{\llap{\raisebox{2pt}{$<$}}}}}}%
\newcommand{\gsim}	{\makebox[15pt]
           {\mbox{{\raisebox{-3pt}{\tsim}}{\llap{\raisebox{2pt}{$>$}}}}}}%
\newcommand\degree   {{\ifmmode^\circ\else$^\circ$\fi}} 

\received{July 10, 2001}
\revised{April 16, 2002}
\accepted{April 18, 2002}
\articleid{2002}{MS~54422}

\shortauthors{R.A.~Windhorst \etal }
\shorttitle{An \HST\ mid-UV survey of nearby galaxies}

\slugcomment{ApJS MS 54422: Revised April 16, 2002 / Accepted April 18, 2002}

\begin{document}

\title {\large\bf An \HST\ Survey of the Mid-UV Morphology of Nearby
        Galaxies\altaffilmark{1,2}\vspace*{4pt}}

\author{Rogier A. Windhorst\altaffilmark{3},
	Violet A. Taylor\altaffilmark{3,13},
	Rolf A. Jansen\altaffilmark{3},
	Stephen C. Odewahn\altaffilmark{3},\\
	Claudia A. T. Chiarenza\altaffilmark{3},\ \
	Christopher J. Conselice\altaffilmark{4,14},\ \
	Richard de Grijs\altaffilmark{5},\\
	Roelof S. de Jong\altaffilmark{6,15},
	John W. MacKenty\altaffilmark{6},
	Paul B. Eskridge\altaffilmark{7,13},
	Jay A. Frogel\altaffilmark{8,13},\\
	John S. Gallagher, iii\altaffilmark{9,14},
	John E. Hibbard\altaffilmark{10,14},
	Lynn D. Matthews\altaffilmark{11,14},\\
	and Robert W. O'Connell\altaffilmark{12}}
\email {Email:\ Rogier.Windhorst@asu.edu}

\altaffiltext{1}{Based on observations made with the NASA/ESA
	\emph{Hubble Space Telescope}, obtained at the Space Telescope
	Science Institute (STScI), which is operated by the Association
	of Universities for Research in Astronomy (AURA), Inc., under
	NASA contract NAS 5-26555}
\altaffiltext{2}{Based in part on observations made with the Vatican
	Advanced Technology Telescope: the Alice P.\ Lennon Telescope
	and the Thomas J.\ Bannan Astrophysics Facility.}


\begin{abstract}

We present a systematic imaging survey of 37 nearby galaxies observed
with the \emph{Hubble Space Telescope} (\HST) Wide Field and Planetary
Camera 2 (WFPC2) in the mid-UV F300W filter, centered at 2930{\AA}, as
well as in the $I$-band (F814W) filter at 8230{\AA}.  Eleven of these
galaxies were also imaged in the F255W filter, centered at 2550{\AA}. 
Our sample is carefully selected to include galaxies of sufficiently
small radius and high predicted mid-UV surface brightness to be
detectable with WFPC2 in one orbit, and covers a wide range of Hubble
types and inclinations.  The mid-UV (2000--3200\AA) spans the gap
between ground-based $UBVR(IJHK)$ images, which are available or were
acquired for the current study, and far-UV images available from the
\Astro/\UIT\ missions for 15 galaxies in our sample.  The first
qualitative results from our study are:\\
(1) Early-type galaxies show a significant decrease in surface
brightness going from the red to the mid-UV, reflecting the absence of a
dominant young stellar population, and in some cases the presence of
significant (central) dust lanes.  Galaxies that are early-types in the
optical show a variety of morphologies in the mid-UV that can lead to a
different morphological classification, although not necessarily as
later--type.  Some early-type galaxies become dominated by a blue
nuclear feature or a point source in the mid-UV, \eg, due to the
presence of a Seyfert nucleus or a LINER.  This is in part due to our
mid-UV surface brightness selection, but it also suggests that part of
the strong apparent evolution of weak AGN in early-type galaxies may be
due to surface brightness dimming of their UV-faint stellar population,
which renders the early-type host galaxies invisible at intermediate to
higher redshifts.\\
(2) About half of the mid-type spiral and star-forming galaxies appear
as a later morphological type in the mid-UV, as \Astro/\UIT\ also found
primarily in the far-UV.  Sometimes these differences are dramatic (\eg,
NGC~6782 shows a spectacular ring of hot stars in the mid-UV).  However,
not all mid-type spiral galaxies look significantly different in the
mid-UV.  Their mid-UV images show a considerable range in the scale and
surface brightness of individual star-forming regions.  Almost without
exception, the mid-type spirals in our sample have their small bulges
bisected by a dust-lane, which often appears to be connected to the
inner spiral arm structure.\\
(3) The majority of the heterogeneous subset of late-type, irregular,
peculiar and merging galaxies display F300W morphologies that are
similar to those seen in F814W, but with important differences due to
recognizable dust-features absorbing the bluer light, and due to hot
stars, star-clusters, and star-formation ``ridges'' that are bright in
the mid-UV.  Less than one third of the galaxies classified as late-type
in the optical appears sufficiently different in the mid-UV to result in
a different classification.\\
Our \HST\ mid-UV survey of nearby galaxies shows that ---when observed
in the rest-frame mid-UV--- early- to mid-type galaxies are more likely
to be misclassified as later types than late-type galaxies are to be
misclassified as earlier types.  This is because the later type galaxies
are dominated by the same young and hot stars in all filters from the
mid-UV to the red, and so have a smaller ``morphological K-correction''
than true earlier type galaxies.  The morphological K-correction can
thus explain part, but certainly not all of the excess faint blue
late-type galaxies seen in deep \HST\ fields. 

\end{abstract}

\keywords{galaxies: elliptical and lenticular --- galaxies: spiral --- 
galaxies: irregular --- galaxies: interactions --- galaxies: peculiar ---
ultraviolet: galaxies}

\flushbottom


\section{\bfseries\sc\large Introduction}
\label{Introduction}

\subsection{The Need for Reliable Classification of Faint Optical Galaxy
	Samples}
\label{TheNeedforReliableClassification}

Since the mid 1990's, faint galaxies have been observed with the
\emph{Hubble Space Telescope} (\HST).  The best statistics, spatial
sampling, and areal coverage are currently achieved in the $I$-band
(F814W).  Good \HST\ morphological information is available today for
\tsim$10^5$ galaxies with $I\lsim 26$ mag.  These galaxies come from the
two Hubble Deep Fields (``HDF''s; Williams \etal 1996, 1998), their
flanking fields, the \HST\ Medium-Deep Survey, and other \HST\ parallel
surveys (\eg, Griffiths \etal 1994; Driver \etal 1995a, 1995b; Odewahn
\etal 1996; Abraham \etal 1996, 1999; Roche \etal 1997; Cohen \etal
2002). 

The median half-light radius of faint field galaxies with $I\sim26$ mag
is $\re\simeq0\farcs 2$--$0\farcs 3$ (Odewahn \etal 1996).  Because of
the high spatial resolution required for proper morphological
classifications (\eg, using artificial neural networks), \HST's 2.4~m
aperture limits the useful wavelength for quantitative galaxy
classification to $4000\lsim\lambda\lsim 8000$\AA.  NICMOS $H$-band
images (\tsim0\farcs 16 FWHM) do not necessarily sufficiently resolve
these faint galaxies to accurately measure their half-light radii and
other important parameters. 

For $I\lsim 23$ mag, a few hundred galaxies have their morphology
measured at \HST\ resolution \emph{and} have measured spectroscopic
redshifts plus velocity dispersion or rotation curves (\eg, Cohen \etal
2000; Lilly \etal 1998, Simard \etal 1999; Vogt \etal 1997), leading to
more quantitative studies of their physical properties such as mass and
$M/L$ ratios.  However, because morphology can be determined with
relative ease with \HST\ for a very large number of galaxies, this
ability allows us to explore a part of parameter space that constrains
galaxy formation and evolution, and that is not yet accessible to
satisfactory spectroscopic study and thus lacks kinematic data for mass
measurements. 

The most dramatic result from these \HST\ morphological studies in the
mid--late 1990's was that at faint fluxes, late-type/irregular galaxies
completely dominate the faint blue galaxy counts (see Driver \etal
1995a, 1995b, 1998; Glazebrook \etal 1995; Odewahn \etal 1996; Windhorst
\etal 1996, 1998b, 2000).  A number of authors interpreted this result
as evidence for the existence of a population of star-forming galaxies
that underwent substantial evolution since $z\lsim 3$ (Ellis \etal 1996;
Pascarelle \etal 1996; Driver \etal 1998; Abraham \etal 1999). 
Ellipticals and early-type spirals, on the other hand, have evolved much
less since $z\lsim 1$ (see Driver \etal 1995a; Lilly \etal 1998; Cohen
\etal 2002).  The photometric redshift distribution as a function of
observed $I$-band morphological type --- N($z_{phot}$, type) ---
suggests a differential evolution of galaxies as a function of type
(Driver \etal 1998), implying a gradual formation of the Hubble sequence
with cosmic time. 

In the deepest \HST\ fields, where morphological classifications are
achievable to $I\lsim$26 mag (\eg, Odewahn \etal 1996; Driver \etal
1998), the sampled redshift range is $z\simeq 1$\,--\,3, while the bulk
of the galaxies is at $z\lsim 2$.  Faint high redshift galaxies observed
in the $I$-band are therefore primarily seen in the rest-frame
mid-ultraviolet (mid-UV) wavelength range, or 2000--3200\AA.\break

\noindent\null\vspace*{-30pt}\leavevmode\makebox[0.485\textwidth]{
\parbox[t]{0.485\textwidth}{\footnotesize
$^{\ 3}$Department of Physics and Astronomy, Arizona State University,
      Box 871504, Tempe, AZ 85287-1504\\
$^{\ 4}$Department of Astronomy, Caltech, MS 105-24 Pasadena, CA 91125\\
$^{\ 5}$Institute of Astronomy, Madingley Road, University of Cambridge,
      Cambridge, CB3 0HA, UK\\
$^{\ 6}$Space Telescope Science Institute, 3700 San Martin Drive,
      Baltimore, MD 21218\\
$^{\ 7}$Department of Physics and Astronomy, Minnesota State University,
      Mankato, MN 56003\\
$^{\ 8}$Code S, NASA Headquarters, 300 D Street SW, Washington, DC 20546\\
$^{\ 9}$Department of Astronomy, University of Wisconsin, 475 N. Charter
      St., Madison, WI 53706-1582\\
$^{10}$National Radio Astronomy Observatory, 520 Edgemont Road,
      Charlottesville, VA 22903-2475\\
$^{11}$Center for Astrophysics, 60 Garden Street, Cambridge, MA 02138\\
$^{12}$Department of Astronomy, University of Virginia, P.O. Box 3818,
      Charlottesville, VA 22903-0818\\
$^{13}$Visiting Astronomer, Cerro Tololo International Observatory,
      National Optical Astronomy Observatory, which is operated by AURA,
      Inc., under cooperative agreement with the National Science
      Foundation (NSF).\\
$^{14}$Visiting Astronomer, Kitt Peak National Observatory, National
      Optical Astronomy Observatory, which is operated by AURA, Inc.,
      under cooperative agreement with the NSF.\\
$^{15}$Hubble Fellow.\\}
}

\rule{0pt}{20pt}

\noindent The reliability of results derived from the faint \HST\ galaxy
morphologies is therefore fundamentally limited by the uncertain
rest-frame mid-UV morphology of nearby galaxies.  For instance, although
these faint, late-type/irregular objects resemble some classes of nearby
late-type and peculiar galaxies (\eg, Hibbard \& Vacca 1997), they need
not be physically late-type objects; instead, they may be earlier-type
galaxies that look dramatically different in the rest-frame UV.  Because
of the wavelength dependence of nearby galaxy morphology (especially
toward the UV), the outcome of faint galaxy classifications will depend
on the rest-frame wavelength sampled.  This ``morphological
K-correction'' can be quite significant, even between the $B$ and
near-IR bands (\eg, Knapen \& Beckman 1996), and must be quantified in
order to distinguish genuine evolutionary effects from simple band-pass
shifting. 

The mid-UV is the optimal wavelength region for comparing the galaxies
seen with \HST\ in the $I$-band at $z\gsim 2$ to nearby samples, and to
address the question of whether the numerous late-type objects that
dominate the faint blue galaxy counts are truly new classes of objects,
or just reflect the redshifted UV morphology of ordinary galaxies as
seen nearby.  In the mid-UV, one currently attains the highest
resolution \emph{plus} largest field-of-view (``FOV'') with the \HST\
Wide Field and Planetary Camera 2 (``WFPC2''). 

In this paper, we therefore present a systematic WFPC2 mid-UV imaging
survey for a representative sample of 37 nearby galaxies.  Having
observations for both nearby and distant objects at the \emph{same}
rest-frame wavelength will allow us to calibrate the dependence of
morphology on rest-frame wavelength and, thus, to improve our ability to
interpret the morphology of distant galaxies.  Such a survey will also
help us better understand the \emph{physical} drivers of the rest-frame
mid-UV emission, \ie, the relation between star formation (SF) and the
global physical characteristics of galaxies, their recent star-formation
history, and the role of dust absorption and scattering.

\subsection {UV Morphology of Nearby Galaxies as Benchmark for High
	Redshift Classifications}
\label{TheRoleofMidtoFarUVMorphology}

Observations of high redshift galaxies in the optical spectral region
detect light that was emitted as UV luminosity in these galaxies'
rest-frames.  They therefore primarily trace high surface brightness
(``SB'') regions populated by high densities of young ($\!\lsim\!1$~Gyr)
hot stellar populations.  The dominance of young stellar complexes in
producing UV light from star-forming galaxies was recognized from early
satellite UV photometric observations (\eg, Code \& Welch 1982; Israel
\etal 1986), {\it IUE} far-UV spectra (\eg, Kinney \etal 1993, and
references therein), as well as imaging of a few key objects from
sounding rockets (\eg, Carruthers \etal 1978; Stecher \etal 1982). 
These results were confirmed when UV images became available from the
FOCA balloon observations (Millard \etal 1992) and the {\it Astro}/\UIT\
flights.  The latter imaged \tsim100 nearby galaxies in the far-UV
(1500\AA) and \tsim40 in the mid-UV (2500\AA) at a resolution of
$\sim$3\arcsec\ FWHM (Bohlin \etal 1991; Hill \etal 1992; Kuchinski
\etal 2000; Marcum \etal 2001).  Morphological K-corrections derived
from these \UIT\ images (and from earlier sounding rocket data) were
first investigated by Bohlin \etal (1991) and Giavalisco \etal (1996). 
This, as well as the FOCA data, showed that galaxy morphology changes
quite dramatically below 3600--4000\AA, where the hot (young) stellar
population --- located mainly in spiral arms and \HII-regions ---
dominates the spectral energy distribution (SED) and where bulges
essentially disappear (see also Burgarella \etal 2001; Kuchinski \etal
2001).  Galaxies therefore often appear to be of later Hubble type the
further one looks into the rest-frame UV.  Qualitatively, this is easy
to understand: in the optical/near-IR, we see the accumulated luminous
phases of long-lived ($>$1~Gyr) stars, which emit most of their energy
at longer wavelengths, whereas the mid-UV samples the star-formation
rate (SFR) averaged over the past Gyr or less.  The mid-UV includes the
longest wavelengths where young stars can dominate the integrated galaxy
light and traces primarily presently active star forming regions, or
those regions where star-formation has only recently shut down. 

Far-UV images of nearby \emph{early-type} galaxies are not always
appropriate for high redshift comparisons; they may be affected by the
``UV upturn'', a spectral feature that is likely caused by hot,
\emph{low-mass, old} stars (O'Connell \etal 1992; O'Connell 1999), and
that only appears at relatively recent epochs ($z\lsim 0.3$; Burstein
\etal 1988; Greggio \& Renzini 1990; Brown \etal 2000).  However,
middle-aged stellar systems can be very dim in the UV compared to either
older or younger objects, as in the case in, \eg, M32.  In galaxies of
all types the far--mid-UV emission can also be modulated by dust.

\subsection{Other Ways of Addressing the Morphological K-correction}
\label{OtherWays}

A number of authors have explored the effects of band-pass shifting
using multicolor optical images to extrapolate to the rest-frame UV on a
pixel-by-pixel basis.  This has been done both using ground-based data
(\eg, Hibbard \& Vacca, 1997; Brinchmann \etal 1998), FOCA mid-UV images
(Burgarella \etal 2001), \UIT\ far-UV images (Kuchinski \etal 2000), and
\HST\ images of galaxies at moderate redshifts (Abraham \etal 1999;
Bouwens, Broadhurst, \& Silk 1998).  The peculiarities in shape and size
distributions measured in deep \HST\ surveys are found to considerably
exceed the effects of band-pass shifting.  While this is a robust
result, these methods may not capture the full range of rest-frame UV
galaxy SEDs, and are less suitable for making detailed comparisons with
the local universe. 

For instance, there is considerable scatter in the mid-UV for galaxies
of a given optical color.  Donas, Milliard \& Laget (1995) find a 3
magnitude range in ($\mbox{UV}-B$) colors at a given $(B-R)$ color in a
faint space-based galaxy sample, where a 2000\AA\ filter was used to
measure the UV flux.  This UV/optical decoupling is confirmed by the
spectroscopy of Treyer \etal (1998) and implies that the true
evolutionary history of galaxies can be very different from that
inferred from optical data alone, and that mid-UV observations are
necessary for a better understanding. 

Recent observations of the HDF-North with \HST/NICMOS show that some
galaxies out to $z\sim2.5$ appear `regular' and fit into the Hubble
sequence in the near-IR, but look more patchy and irregular at shorter
wavelengths (Dickinson \etal 2001).  Galaxies that look disturbed in the
optical, on the other hand, in general also look very similar in the
rest-frame UV.  These distorted high-$z$ galaxies also tend to be high
SB objects in the UV, forming stars at a high rate.  Nearby galaxies
whose structure is dominated by star-formation also have similar
morphologies in the rest-frame optical and mid-UV (\eg, Conselice \etal
2000), which suggests that some nearby star-bursts galaxies may be
analogs to these `peculiar' high-$z$ objects.

\subsection{Goals of our \HST\ Mid-UV Imaging Survey}
\label{ScienceGoals}

In this section we briefly summarize the main science goals of our
project, which determine the size and properties of our WFPC2 mid-UV
sample.  We aim to:\\[-12pt]

\noindent (1) provide local benchmark images that we can artificially
redshift to $z\simeq 1$--3 for quantitative comparison to the morphology
and photometric properties of high redshift galaxies;\\[-12pt]

\noindent (2) consistently classify polychromatic structures within
galaxies from 0.2--2\micron\ with artificial neural networks.\\[-12pt]

\noindent (3) map the spatial distribution, luminosities, and sizes of
star-forming regions dominating the UV morphology, and relate these to
global galaxy properties;\\[-12pt]

\noindent (4) map the effects of dust at mid-UV wavelengths in galaxies
as a function of type, inclination, and galactocentric radius;\\[-12pt]

\noindent In the present paper, we present the benchmark mid-UV images
and a qualitative discussion of our results in the context of these
science goals.  More quantitative studies will be presented in
subsequent papers by members of our mid-UV team. 

This paper is organized as follows.  In section
\ref{WFPC2SampleSelectionandWFPC2Observations}, we outline our survey
strategy, describe the sample selection and discuss the WFPC2
observations.  We also discuss the data processing and calibration.  Our
WFPC2 mid-UV and ground-based optical images are presented in
\S~\ref{WFPC2MidUV-OpticalATLAS}, where we also discuss individual
objects.  We describe the general trends in galaxy morphology as
observed from the mid-UV through the optical in
\S~\ref{DiscussionofGeneralTrends}.  And in \S~\ref{Conclusions}, we
present our main conclusions from this data set.


\section{\bfseries\sc\large Strategy, Sample Selection and Observations}
\label{WFPC2SampleSelectionandWFPC2Observations} 

\subsection{WFPC2 Mid-UV Survey Strategy}
\label{Strategy} 

\subsubsection{Mid-UV Filters}
\label{Strategy-Filters}

With the \HST\ Wide Field and Planetary Camera 2 (``WFPC2''), we have
obtained images of 37 nearby galaxies through one or, whenever possible,
two wide-band mid-UV filters below the atmospheric cutoff.  These
filters are F300W ($\lambda_{cent}\simeq 2930$\AA; $\Delta\lambda\simeq
740$\AA\ FWHM) and F255W ($\lambda_{cent}\simeq 2550$\AA;
$\Delta\lambda\simeq 395$\AA\ FWHM), which provide reasonable red-leak
suppression.  The F255W, F300W, and the Johnson $U$ and $B$ filters are
approximately equally spaced in energy (\ie, in the logarithm of the
wavelength), and so add significantly to the existing ground-based
optical--near-IR color baseline. 

Since the \HST/WFPC2 system throughput is \tsim2.0\% in F300W and
\tsim0.5\% in F255W (Biretta \etal 2001; Appendix 1), we can only detect
the highest SB, bluest objects in F255W in a single \HST\ orbit and so
have selected our sample accordingly.  The mid-UV is the longest
wavelength where younger stars can dominate the integrated galaxy light,
and therefore the regime of choice to measure the SFR averaged over
$\lsim1$~Gyr.  We have observed \emph{all} selected galaxies through
the F300W filter, spending no more than one full orbit per galaxy.  In
that same orbit a short exposure through a red filter (F814W) is taken
for adequate red-leak correction (see \S\ref{Redleak}).  For galaxies in
the \HST\ Continuous Viewing Zone (CVZ), we also took exposures in the
F255W filter (see \S\ref{Strategy-CVZ}).

\subsubsection{Predicted Mid-UV Surface Brightness} 
\label{Strategy-PredictedSB}

We predict the average mid-UV SB, \SB, for a given galaxy from its total
$B$ magnitude, $B_T$, its $(U-B)$ color, its half-light radius, \re, and
its ellipticity, $b/a$, (as tabulated in or derived from the RC3
catalogue [de~Vaucouleurs \etal 1991], or the NASA/IPAC Extragalactic
Database, NED) as following:

\begin{eqnarray}
\SB\ = \mbox{\sl F300W}_T + 0.75 + 2.5\,\log(\pi\; r_e^2\times b/a)\ ,
\end{eqnarray}

\noindent \ie, half the total predicted F300W magnitude,
\textsl{F300W}$_T$, within the effective area.  We used the updated
Bruzual \& Charlot (1993) models to transform the $(U-B)$ color for each
galaxy type to a predicted (\textsl{F300W}$-B$) color, from which
\textsl{F300W}$_T$ follows.  To a reasonable approximation we find
(\textsl{F300W}$-B) \simeq 2\times(U-B)$.  In the absence of a $(U-B)$
color, a prediction for $(U-B)$ was made from the measured $(B-V)$ color
and the known $(U-B)$ vs.\ $(B-V)$ relation for RC3 galaxies as a
function of galaxy type (de~Vaucouleurs \etal 1991).  Our sample was
selected to have $18\lsim \SB\lsim 22.5$--23.0 \magarc.  For this range
in SB, a galaxy can be detected out to $r\simeq 2$--$3\, \re$ with WFPC2
in one orbit with sufficiently high S/N to allow morphological features
to be recognized. 

The bias toward selecting higher SB galaxies can be addressed as in
Driver \etal (1995b).  In short, selecting high SB galaxies as the
nearby template objects is not an overriding concern, since the
high-redshift samples are similarly biased (or more so) in favor of high
SB galaxies due to the severe cosmological SB dimming.  For
monochromatic light, SB-dimming is proportional to $(1+z)^{(4+\alpha)}$,
with $\alpha$ the spectral index if the object spectrum were to be
represented by a power-law SED. 

The resolution of \HST's Optical Telescope Assembly (OTA) in F300W is
\tsim$0\farcs 04$.  This is somewhat larger than \HST's formal
diffraction limit at 2930\AA\ ($1.22\cdot\lambda/\mbox{D}\simeq 0\farcs
03$), which doesn't set in until longward of 4000\AA\ due to mirror
micro-roughness.  The WFPC2 WFC pixel size is $\simeq0\farcs
0996$/pixel.  Hence, because the UV images are already severely
undersampled, on-chip rebinning to gain SB-sensitivity is not an option. 
Instead, where needed, we can rebin the images in the post-processing
stage to measure the outskirts to fainter SB-levels.  This improves the
SB-sensitivity (see \S~\ref{SNAchieved}) in the outskirts to
\tsim25.8--26.3 \magarc\ in F300W and to \tsim23.8--24.2 \magarc\ in
F255W, sufficient to get good light-profiles for $r\lsim 2$--3\,\re.

\subsubsection{Target Size and Placement Inside WFPC2}
\label{Strategy-SizePlacement}

The WFPC2 FOV measures \tsim2.5\arcmin\ (along the WFC CCDs).  We
selected the sample to fit within the FOV, and preferably within a
single $75\arcsec\times75\arcsec$ WFC CCD, allowing us to derive
reliable SB-profiles without having to mosaic multiple WFPC2 fields. 
For galaxies with a $B$-band half-light radius in the range
$0\farcm1\lsim\re\lsim1\farcm0$ (as derived from the RC3 catalogue,
de~Vaucouleurs \etal 1991), about \tsim3--5 scale-lengths fit in a
single WFPC2 field. 

For most of our sample galaxies, the nucleus has been placed on WFPC2's
WF3 CCD, near pixel $(X,Y)=(300,300)$, so that both the WF2 and WF4 CCDs
maximally sample the galaxies' outskirts, allowing optimal subtraction
of any sky-background when mosaicing the four WFPC2 CCDs.  For some of
the larger galaxies and for galaxies in pairs or small groups, we center
the object(s) in another part of the WFPC2 FOV, or constrain the \HST\
roll angle (``ORIENT'') to assure that the largest possible portion of
the galaxy or galaxy group is observed.

\subsubsection{The \HST\ Continuous Viewing Zone}
\label{Strategy-CVZ}

Part of the galaxy sample is located in the \HST\ Continuous Viewing
Zone (CVZ), \ie, at $53\arcdeg\le|\mbox{DEC}|\le 72\arcdeg$, where
objects are observable for an entire \HST\ orbit, typically doubling the
available integration time.  For any such galaxies, we were able to
obtain F255W as well as F300W images without the cost of an extra \HST\
orbit.  We select the sample to maximize the fraction of galaxies in the
CVZ. 

Since the Zodiacal background reflects the color of the Sun, the
sky-background will be darker in the F255W filter (\tsim24.5--25.0
\magarc) than in the F300W filter (\tsim24.0 \magarc), partly
compensating for the lower sensitivity in F255W.  Observations in the
CVZ may suffer from higher sky-background levels due to the Earth's limb
(Williams \etal 1996).  We minimize the probability of excessive
sky-levels by interspersing the exposures in the different filters using
the sequence: F814W, F255W, F300W, F255W, F300W, F814W, F255W, and
F300W.  This sequence ensures that never more than one F255W or F300W
exposure is taken close to the Earth's limb in a full CVZ orbit, and
also minimizes the number of Fine Guidance Sensor (FGS) motions needed
to create a pointing dither-pattern.  For non-CVZ targets we use the
sequence: F814W, F300W, F300W, and F814W, to push the F300W exposures
farthest from the Earth's limb.

\subsection{Sample Selection}

\subsubsection{Available Ground- and Space-based Observations}
\label{AvailableSpaceBased}

In selecting our galaxy sample we aimed to take advantage of as much
pre-existing ground- and space-based UV, optical and near-IR images as
possible.  In particular, high priority was given to galaxies that have
been observed with \UIT\ in the far-UV (1500\AA) and obeyed the SB and
size criteria (\S~\ref{Strategy-PredictedSB} and
\S~\ref{Strategy-SizePlacement}).  Some of these galaxies are close to
the \UIT\ detection limit, as they are minor members in galaxy groups
targeted by \UIT. 

Also, nearby galaxies already observed with \HST/WFPC2 in the mid-UV
F300W and/or F255W filter that match our criteria will be included in
our final sample.  When the final Cycle 9 target list was compiled in
early 2000, the \HST\ Archive contained 15 such galaxies with F300W
images obtained and two more imaged in F255W.  These Archival
observations came primarily from Cycle~4 program GO~5381 (PI: Koratkar)
and Cycle~6 program GO~6355 (PI: Giavalisco).  We will analyze these 17
Archival mid-UV images in a separate paper (Taylor \etal 2002, in
preparation).  The current Cycle~9 sample is designed and must be
understood to be complementary to this existing Archival sample. 

\label{AvailableGroundBased}

Over the last decade, several of us have systematically imaged from the
ground a total of about 750 nearby galaxies covering all Hubble types
and inclinations.  The majority of these galaxies --- which includes
most of the galaxies imaged in the far-UV with \UIT\ (primarily at
1500\AA) --- were observed in $UBVRI$, but a good fraction was observed
in the near-IR $JHK$ filters as well (Eskridge \etal 2002b).  The data
base thus compiled consists of seven individual galaxy samples:

\noindent\textbf{(1)} 86 face-on spiral galaxies of de~Jong \& 
van~der Kruit (1994);

\noindent\textbf{(2)} 220 galaxies with Hubble types later than S0 from
the OSU $BVRJHK$ survey (Frogel \etal 1996; Eskridge \etal 2000, 2002b)
(all inclinations);

\noindent\textbf{(3)} 113 galaxies from the Frei \etal (1996) survey in
$B$ and $R$, to which we added $U$ and $V$-band images (good number of
early-type galaxies; all inclinations; Taylor \etal 2002; in prep.);

\noindent\textbf{(4)} 100 galaxies with 1500\AA\ and 40 with 2500\AA\
images from the \Astro/\UIT\ mission (Kuchinski \etal 2000; Marcum \etal
2001) (all inclinations);

\noindent\textbf{(5)} 48 edge-on galaxies (de~Grijs \& Peletier 1997;
de~Grijs, Peletier, \& van~der~Kruit, 1997);

\noindent\textbf{(6)} 150 late-type UGC~galaxies selected to be
morphologically irregular, peculiar, or merging (Hibbard \& Vacca 1997;
Hibbard \etal 2002).  These have a range of impact parameters (\ie, the
distance between the two interacting or colliding objects); and

\noindent\textbf{(7)} 49 late-type dwarf spiral galaxies (Matthews \&
Gallagher 1997) and compact, high-SB luminous blue galaxies (Gallagher
\etal 1989, 2001) (all inclinations). 

\noindent with some overlap among the samples.  The last two sets were
specifically added to provide a good number of likely local counterparts
of the peculiar and irregular galaxies seen with \HST\ in large numbers
at high redshifts.  The $UBVRI$ + $JHK$ images from these ground-based
surveys, where available, are included in the data presented in
\S~\ref{WFPC2MidUV-OpticalATLAS} (Fig.~4).

\subsubsection{Required Statistics}
\label{RequiredStatistics}

To date, the \HST\ Archive contains over 10,000 orbits of data on a
multitude of distant galaxy projects (including parallels).  To match
these numerous observations of distant galaxies, we need a statistically
significant and representative comparison sample of nearby galaxies that
includes all morphological types.  To cover a broad range of nearby
galaxies, we will use three broad bins of Hubble types: (1) early-type
galaxies (E--S0), (2) early-mid type spiral galaxies (Sa--Sc), and (3)
late-type galaxies, irregulars (Sd--Irr), plus peculiars and mergers. 
For statistical studies as a function of photometric parameters, we will
cover a range in inclination angles in three bins of $\sin{i}$: edge-on,
face-on, and intermediate angles.  Thus, we need about 54 galaxies in
total, to populate each of the nine type--inclination bins with 6
galaxies to begin to average over galaxy properties. 

Of the 20 galaxies with WFPC2 F300W images in the Archive as of early
2000 (mostly E/S0/Sabc's), about 17 are small enough to be contained
within the WFPC2 FOV.  To this available mid-UV sample we add here a
complementary set of 37 galaxies, which we selected from the
ground-based samples of about 750 galaxies described in section
\ref{AvailableGroundBased}.  These objects comprise 24 new irregulars,
peculiars, and merging/interacting galaxies, and 13 new E/S0/Sabc's to
complement the 17 usable galaxies available in the \HST\ Archive.

\subsubsection{Selection of the \HST/WFPC2 Mid-UV Sample}
\label{HSTWFPC2SampleSelection}

The selection criteria for inclusion in the \HST/WFPC2 sample of 37
galaxies are:

\noindent\textbf{(a)} the $B$-band half-light radius should fall within
the range $0\farcm 1\lsim\re\lsim 1\farcm 0$;

\noindent\textbf{(b)} the predicted average SB in F300W out to
$r\simeq\re$ should be in the range $18\lsim \SB\lsim 22.5$--23.0
\magarc;

\noindent\textbf{(c)} the distribution of Hubble types should be
representative for nearby galaxy samples when combined with the set of
archival galaxies;

\noindent\textbf{(d)} the distribution over apparent axis ratio (and
hence inclination) should resemble that of nearby galaxy samples;

\noindent\textbf{(e)} priority should be given, first, to the 15
galaxies with available \UIT\ 1500\AA\ images that obey the other
selection criteria above, and then to galaxies with multi-band
ground-based images available;

\noindent\textbf{(f)} as much as possible, select galaxies that are
schedulable in the \HST\ CVZ, while satisfying \textbf{(a)} --
\textbf{(e)}.  Replace candidates that are difficult to schedule by
alternates. 

In Table~1 we list the properties of the selected sample.  Most of the
table entries are taken or derived from values listed in the RC3
catalogue (de~Vaucouleurs \etal 1991).  Cols.~2 and 3 list the
coordinates of the galaxy centers as measured on the Digital Sky Survey
(DSS).  In a few cases of complex galaxies, merging/interacting
galaxies, or galaxies in small groups, the listed coordinates refer to a
``center of mass'' position, allowing optimal centering inside the WFPC2
FOV.  Col.~4 shows the apparent axis ratio of the minor and major axis
isophotal diameter measured at the $\mu_B=25.0$ \magarc\ isophote.  The
half-light radius, \re\ and the radius at the $\mu_B=25.0$ \magarc\
isophote, $r_{25}$, are listed in Cols.~5 and 6, respectively; the
predicted average SB in F300W out to \re\ and $r_{25}$, and the
predicted total \textsl{F300W} magnitude (see section 2.2) are tabulated
in Cols.~7, 8, and 12.  Col.~13 lists the revised T-type (numeric
morphological type on the 16-step de Vaucouleurs scale).  The redshifts
in Col.~14 are given with respect to the Galactic Standard of Rest. 
Col.~15 lists for each galaxy the physical image scale in pc/(\arcsec)
at the distance of the galaxy, assuming a Hubble constant
$H_0=67$~\kmsMpc.  And Col.~16 gives the physical size of a single WFC
CCD-frame at that redshift.  For the typical redshift range,
$z\simeq0.004$---0.02 (\ie, Virgo--Coma cluster like distances), a
single WFPC2 CCD covers 6.8--34 kpc. 

The SB-selection of the sample is illustrated in Fig.~1, which shows the
normalized distributions of the predicted average \SB\ out to \re\ for
the full (54 galaxy) sample and for the 37 galaxies in the Cycle~9 WFPC2
mid-UV sample, and compares these with the distribution over SB derived
for the 3009 galaxies in the RC3 with measured $B_T$, $(U-B)$ and \re. 
Clearly, we sample predominantly galaxies with higher SB than the median
SB-value in the RC3, which is $\SB\simeq 22.3$ \magarc. 

As nearby galaxy samples like the RC3 are also limited by significant
SB-selection (Disney 1976, Driver 1999), this thus means that our Cycle
9 \HST\ sample represents only the tip of the local galaxy
SB-distribution.  Fig.~1 shows that our sample covers only about the
brightest 50\% SB-percentile seen in the RC3, while it over-represents
the highest SB galaxies seen in the RC3.  If one wanted to use our
sample to represent all galaxies in the RC3 with $\SB\le 23$ \magarc,
one would have to apply weights to each of our galaxies in a given
observed SB-bin that could be derived from dividing the two
distributions over SB in Fig.~1.  This will obviously not work for
galaxies with $\SB\ge 23$ \magarc, since almost no galaxies at average
SB-levels that faint were observed with \HST. 

As aforementioned, this SB-selection is justified when the primary goal
is to construct a sample for comparison of nearby and high-redshift
galaxies, since the strong cosmological SB-dimming limits our view of
the high-redshift universe in a similar way (or worse).  For example,
simulations by Windhorst \etal (2002), who investigate how the galaxies
in the present sample would appear to NGST under realistic observing
conditions, show that at very high redshifts ($z\sim7$--15) only the
highest SB mid-UV structures (with average rest-frame \SB\ out to \re\
brighter than 20--21 \magarc) will be visible to NGST. 

Figures~2\emph{a}--\emph{d} show the distributions of morphological
type, of apparent axis ratio $b/a$, of the estimated average $(U-B)$
color within the effective radius, and of the effective radius \re,
within the selected sample.  These figures will help the reader to
evaluate the usefulness or the weight to apply to individual galaxies in
the present sample when using these data for his or her particular
purpose.  We stress that the present sample is \emph{not formally
complete} and, therefore is not suitable for studies that rely on
complete statistics (\ie, studies that cover the complete local galaxy
luminosity function or the complete range in local galaxy SB). 
Nonetheless, important trends are visible in Figs.~2\emph{a}--2\emph{d}
that will help the reader to make our sample useful for high-redshift
comparisons.  These trends are:

\noindent $\bullet$ Fig.~2\emph{a} shows that our selected mid-UV
galaxies overemphasize late-types a bit more than the RC3.  This was
done on purpose, since late-types are the dominant galaxy population at
high redshifts (see Driver \etal 1995a). 

\noindent $\bullet$ Fig.~2\emph{b} shows that our selected mid-UV
galaxies sample the $b/a$ distribution of the RC3 fairly, given the
small number statistics.  Since Odewahn, Burstein, \& Windhorst (1997)
found no significant trend in the galaxy $b/a$ distribution from the RC3
level ($B\lsim 15$ mag) all the way down to the HDF limit ($B\lsim 28$
mag), our nearby comparison sample is thus a fair one in terms of galaxy
ellipticities at all redshifts. 

\noindent $\bullet$ Fig.~2\emph{c} shows that our selected mid-UV
galaxies overemphasize the bluest object in $(U-B)$ a bit more than the
RC3.  Like Fig.~2\emph{a}, this was done on purpose, since faint blue
galaxies are the dominant population at faint magnitudes. 

\noindent $\bullet$ Fig.~2\emph{d} shows that our selected mid-UV
galaxies overemphasize the intrinsically smaller galaxies in $r_e$ more
than the RC3, which was again done on purpose, since faint blue galaxies
are on average significantly smaller than the ones seen nearby (see
Odewahn \etal 1996, Pascarelle \etal 1996, Cohen \etal 2002). 

Again, any non-representativeness of our galaxy sample compared to the
RC3, or to any other galaxy catalog for that matter, can be addressed by
applying the appropriate weights to individual galaxies in our sample
when comparing it to any other samples.  These weights can be determined
from the ratios between the observed distributions in
Fig.~2\emph{a}--2\emph{d}, or between our sample and any other sample
than the RC3.

\subsection{Observations}
\label{Observations}

\subsubsection{\HST/WFPC2 Observations}
\label{WFPC2cycle9-Observations}

All 37 selected galaxies were observed with WFPC2 during Cycle~9. 
Typically, we exposed 2$\times$800--1000~s in F300W and
2$\times$100--160~s in F814W.  Exposures times were made flexible to
optimally use the full \HST\ orbit allocated per galaxy.  Of the 24
galaxies located in the CVZ, 11 could indeed be scheduled there during
Cycle 9.  For these, exposure times were 3$\times$400--500~s in both the
F255W and F300W filters, and 2$\times$100--160~s in F814W.  In none of
the 37 Cycle 9 targets did we see evidence for elevated sky-levels in
F300W (see \S~\ref{Strategy-CVZ}). 

We adopted a linear dither pattern such that exposures in the same
filter are offset by $n+0.34$ WFC pixels in the case of two (non-CVZ),
and $n+0.33$ and $n+0.66$ WFC pixels in the case of three exposures per
filter (CVZ).  Here, $n$ is a multiple of 5 pixels (ideally 10--30
pixels in both directions to fill in the vignetted regions between the
WFPC2 CCD's), and the pixel fractions assure optimal drizzling in a
later stage.  The actual shifts may slightly deviate from the listed
values depending on the quality of the FGS lock achieved for a
particular galaxy.

\subsubsection{Ground-based Observations}
\label{GroundBasedObservations}

Most of the ground-based $UBVRI$ CCD images for the 37 galaxies in the
present sample were obtained with the 1.8~m Vatican Advanced Technology
Telescope (VATT) on Mt.~Graham, with additional images taken with the
0.9~m telescope at Cerro Tololo (CTIO), the Danish 1.5~m telescope at
the European Southern Observatory (ESO), the 2.1~m telescope at Kitt
Peak (KPNO), the 2.2~m University of Hawaii telescope at Mauna Kea (UH),
or the 1.0~m Jacobus Kapteyn Telescope (JKT) at La Palma.  The optical
data were obtained mostly with 2k$\times$2k CCD's.  Further details are
given in Matthews \& Gallagher (1997), Eskridge \etal (2002b), and
Taylor \etal (2002, in prep.). 

Ground-based near-IR imaging was done using the 1.5~m telescope at Cerro
Tololo, the ESO~2.2~m telescope, the Carnegie 2.5~m du~Pont telescope at
Las Campanas (LCO), or the 3.8~m United Kingdom InfraRed Telescope
(UKIRT) at Mauna Kea, mostly with 256$^2$ HgCdTe-arrays (and 1024$^2$ in
the case of the du~Pont telescope).  Images were obtained in $JHK$, or a
subset of these filters.  For further details on these near-IR data we
refer to de~Grijs \etal (1997), de~Jong \& van~der Kruit (1994), and
Eskridge \etal (2002b). 

The $JHK$ filters sample the underlying cooler (and usually older)
stellar population, which in general also dominates the $R$ and
$I$-bands, so that the $RIJHK$ filters in general reveal similar galaxy
morphology.  Morphological contrast is added through the younger stellar
populations in the bluer and UV filters.  Since the older stellar
population dominates the $RIJHK$ fluxes of earlier type galaxies, a
comparison of these filters can in principle provide a measure of the
dust content, whose effects are strongly wavelength dependent --- noting
the usual caveat that effects from dust and metallicity are hard to
separate, and require spatially resolved spectroscopy in addition to
broad-band photometry to address this quantitatively. 

We aimed for a photometric zero-point accuracy better than 0.03--0.05
mag, in the optical as well as the near-IR.  The ground-based CCD images
were reduced following standard procedures, as described in Matthews \&
Gallagher (1997), and Neuschaefer \& Windhorst (1995). 

In Table~2 we summarize our Cycle~9 \HST\ observations together with the
\UIT\ and ground-based images available to date.  For WFPC2
observations, we list the \HST\ roll-angle (\texttt{PA\_V3}) and the
nominal FWHM of the \HST\ PSF; for ground-based observation, we list the
airmass and the median FWHM in each image stack.  These FWHM values were
measured from unsaturated stellar images using \texttt{SExtractor}
(Bertin \& Arnouts, 1996).  We clipped the FWHM distribution at
$0\farcs3$ to filter out cosmic rays and bad pixels, and at 5\arcsec\ to
filter out misclassified extended objects (\eg, \HII\ or star-forming
regions inside the galaxies of interest).  These values, and the FWHM
values of Moffat (1969) profiles as fitted using the task
\texttt{imexamine} within IRAF, are consistent to within 2.5\% with a
dispersion of 9\%.  FWHM values could be measured for 124 (out of 131)
ground-based images.  For the remaining images that did not contain any
unsaturated stars, the FWHM was estimated from the images taken closest
in time that did.  The mean (median) seeing over all ground-based images
is $2\farcs04\pm0\farcs52$ ($2\farcs00\pm0\farcs52$) (1-$\sigma$).  This
rather large average image-FWHM is due to the majority of VATT images
having a focus component in their PSF as a result of the rapid change of
focus with temperature in this f/1 telescope (for details, see Taylor
\etal 2002).  Even with FWHM \tsim\ 2\farcs0, these ground-based images
are quite suitable for our current purpose of galaxy surface photometry
in $UBVR$.

\subsubsection{Archival \HST\ Images in Additional Filters}
\label{ArchivalHSTImages}

Several of the galaxies in the present sample have been observed with
\HST\ in filters other than F300W and F255W, or with instruments other
than WFPC2.  Maoz \etal (1996), for example, imaged the central
$22\arcsec\times22\arcsec$ in 110 nearby galaxies in the mid-UV
(2300\AA) using the Faint Object Camera (FOC).  The red-leak in the FOC
mid-UV filters, however, is more severe (\tsim24\%; Keel \& Windhorst
1993) than that in the WFPC2 filters ($\simeq5$--7\%; see
\S\ref{Redleak}), and the FOV is much smaller.  Sixteen of our objects
have been observed in other WFPC2 filters before --- generally in a
heterogeneous mixture of filters: F170W (far-UV), F336W, F439W, F555W,
F606W, F702W, and/or some additional F814W images.  These data are
summarized in Table~3.  In the archival WFPC2 exposures, the galaxy
nucleus was often placed in the Planetary Camera (PC), which due to its
smaller pixel size is less suited for accurate surface-photometry. 
Also, for galaxies \tsim1\arcmin\ in extent, more of a galaxy tends to
be lost outside WFPC2's FOV when its nucleus is placed in the PC.  We
will study these Archival WFPC2 images in the other filters, where
appropriate, in future papers. 

Where Archival NICMOS images in the F110M/F110W, F160W or F222M filters
are available (GTO~7218, PI: Rieke, see \eg, Alonso-Herrero \etal 2000;
GTO~7219, PI: Scoville, see Scoville \etal 2000; GO~7268, PI:
van~der~Marel, see Ravindranath \etal 2001; SNAP~7328, PI: Malkan, see
Quillen \etal 2001; SNAP~7330, PI: Mulchaey, see Regan \& Mulchaey 1999)
we use them in lieu of ground-based $JHK$ data in Table~2 and Fig.~4.

\subsection{Calibration and Processing of the \HST\ Observations}
\label{CalibrationandProcessing}

\subsubsection{Standard Processing}
\label{Standard Processing} 

Standard WFPC2 CCD processing following Windhorst \etal (1994b, 1994c,
1998a), Driver \etal (1995a), Neuschaefer \& Windhorst (1995), and
Odewahn \etal (1996) included bias and dark-subtraction, and
flat-fielding.  Photometric calibration was done using the STSDAS
On-The-Fly-Reduction (OTFR) routines as available since summer 2000. 
Custom calibration, in general, does not significantly improve upon the
STScI WFPC2 pipeline, owing to the significant work that went into
building and improving that pipeline.  The OTFR takes into account the
latest improvements in knowledge of the instrument every time one
retrieves data from the \HST\ Archive. 

Because the mid-UV images have extremely low sky-background levels, the
background subtraction is limited by the quality of the bias and dark
current removal.  It is therefore important that the very best possible
biases and latest dark-current and hot-pixel maps are used.  We paid
close attention to whether the correct dark-frames were used when
observations were taken near the monthly warm-up of WFPC2 (to
decontaminate the optics and anneal many of the new hot-pixels).  The
OTFR uses the best available super-dark taken after the relevant science
images, but before the next decontamination.  Hence, it is conservative
in nature --- repairing more pixels than needed, but never too few.  We
re-ran the OTFR on all data one month after the last images for this
project were taken (in April 2001) to incorporate the latest knowledge
on the WFPC2 data.  The difference between this second run and the first
was very small, but went in the direction of removing a more appropriate
(smaller) number of pixels deemed ``hot''. 

We co-added all images in the same filter after registration using
integer pixel shifts.  Our in-house IDL routine \texttt{STCombine}
(Pascarelle \etal 1998; Cohen \etal 2002) was used to optimally remove
the signal induced by the many Cosmic-Ray (CR) hits.  This routine,
optimized for the low signal domain, applies a one-sided 2-$\sigma$
rejection in creating the final stacked image (following Windhorst \etal
1994a).

\subsubsection{Achieved S/N in the WFPC2 Images}
\label{SNAchieved}

The Zodiacal sky-background at the North Ecliptic Pole is \tsim24.0
\magarc\ in F300W (Windhorst \etal 1994b, 1998a) and \tsim24.7 \magarc\
in F255W (Cornett \etal 1994), and is as low in the sunlit part of the
orbit as it is in the occulted part (unless the angle to the Earth's
limb becomes small).  Since our WFPC2 mid-UV images are read-noise
limited, the resulting 1-orbit 1-$\sigma$ SB-sensitivity is
$25.1\pm0.15$ \magarc\ in F300W and $23.0\pm0.15$ \magarc\ in F255W on a
per pixel basis.  The relation between the detected SB-level and the S/N
in a pixel is given by: $\mbox{S/N}=10^{-0.4\,(\mu_{\mbox{\tiny\sl
F300W}}-25.1)}\,$ and $\mbox{S/N}=10^{-0.4\,(\mu_{\mbox{\tiny\sl
F255W}}-23.0)}\,$ for F300W and F255W, respectively.  These mid-UV
SB-limits are consistent with the values expected from the Cycle 5--6
images in F410M and F450W (Pascarelle \etal 1996, Odewahn \etal 1996,
Windhorst \etal 1998a) and the relatively red color of the zodiacal
sky-background.  Taking into the account the $(1+z)^4$ SB-dimming and
the typical $(U-I)$ color of galaxies at $z\simeq 1$--2, the
SB-sensitivity reached in the present data matches that achieved for
typical faint $I\lsim 26$ galaxies seen in deep \HST\ images. 

The 1-orbit 3-$\sigma$ point source sensitivity is $26.4\pm0.15$ mag in
F300W and \tsim24.5 mag in F255W.  Hence, many of the galaxies in our
sample are resolved into their brightest star-forming regions and ---
most likely --- into their OB associations and young star clusters. 
This is not true, however, for most of the merging/interacting galaxies
that were selected into our sample from the sample of Hibbard \etal
(2002).  Because these systems are relatively rare, they tend to be 2--3
times more distant than the bulk of our sample (see Table~1, Col.~14),
and so are not resolved into stars.

\subsubsection{Red-leak}
\label{Redleak}

The F300W filter has a significant red-leak, causing a fraction of an
object's flux longward of 4000\AA\ to be detected in this mid-UV filter. 
Fig.~3.10 of the WFPC2 Handbook (Biretta \etal 2001) shows that red-leak
portion of the $QE\times T$ curve of the F300W filter resembles the
throughput curve of the F814W filter which transmits mostly photons in
the 7000--9000\AA\ range.  Table~3.13 of the WFPC2 handbook suggests
that the red-leak is generally no more than 5\% of the total F300W flux
for stellar populations dominated by stars of spectral type K3V or
earlier, although it can be as much as 10--50\% of the total F300W flux
for stellar populations dominated by M0--M8V stars.  Hence, even for
elliptical galaxies with K-star spectra, the red-leak is expected to be
relatively small, and for late-type galaxies dominated by young hot
stellar populations it should be almost negligible. 

For realistic galaxy SEDs, Eskridge \etal (2002a) find that the red-leak
is typically 5--7\% of the total F300W flux, and never exceeds 10\% of
the F300W flux, not even in the reddest galaxy bulges.  We verified this
for several red galaxies in our sample by subtracting a fraction of the
F814W images from the F300W images, after appropriately rescaling with
the relative exposure time, and making sure both images were registered
the same way.  This fraction of the subtracted F814W image amounted to
7\% of the total F300W flux in the brightest region of the galaxy bulge
that is presumed to be dominated by G8-K3 stars (following the red-leak
as modeled by Eskridge \etal 2002a). 

We found that for the redder stellar populations in those images no
noticeable additional structure was introduced in our F300W images at
the locations of the brightest F814W flux.  To illustrate this, a very
red star is seen just above the center of both edge-on galaxies
ESO033-G22 and IC~4394 in the F814W images of Fig.  3.19 and 3.20.  For
IC~4394, Fig.  4.20 shows how red this star is, where it is seen just
South of the galaxy center.  These stars are saturated in the F814W
images of both galaxies, and at the corresponding locations in the
(non-red-leak corrected) F300W images of Fig.  3.19--3.20, only a very
faint red-leak flux is seen.  These worst case examples show that the
apparent F300W morphology of any of our galaxies would not be
significantly affected by the red-leak in a few of the very reddest and
brightest galaxy areas in the F814W images.  Such areas would have to be
significantly saturated in our F814W exposures to generate significant
red-leak in the F300W images, and none of our targeted galaxies were
saturated anywhere in the F814W images.  Hence, for the current
qualitative presentation of the mid-UV images, and given that our sample
is biased toward the bluer galaxies, we have thus not corrected the
images presented in the mid-UV atlas of \S~\ref{WFPC2MidUV-OpticalATLAS}
for the small contamination by red-leak in the brightest and reddest
areas. 

For accurate quantitative measurements of galaxy properties we will
subtract the red-leak in future papers where necessary (\eg, Eskridge
\etal 2002a).  This will be done together with the determination of
pixel-to-pixel SEDs for each galaxy, in case there are subtle
dependencies of the red-leak on the red SED.  However, to first order
there should be no such dependency, since galaxies of all spectral types
show remarkably little change in their relative SED's between
7000-9000{\AA}.

\subsubsection{Data Archiving}
\label{DataArchiving}

As we expect these data to be of use beyond the scope of our immediate
science goals, we will make all images available to the community in
digital form when this paper goes to press.  We do this, even though the
photometric zero-points for part of the ground-based images are not yet
established.  We will update the FITS headers as new photometry becomes
available (see also Taylor \etal 2002).  Hence, the FITS headers in the
public data base will override any values currently listed in Table~2. 
The images will be made public via ADS and also at the following public
Web-Site at ASU:

{\footnotesize\sf\
\indent\url{www.public.asu.edu/$\sim$caburg/hstuband/newhstuband/}}\\[-8pt]


\noindent Both the \HST\ and ground-based data will be made publicly
available at this site.  The raw WFPC2 data can be obtained from the
\HST\ Archive.  Additional information regarding this survey and its
planning, the observations, and reduction procedures can be found at:

{\footnotesize\sf
\indent\url{www.public.asu.edu/$\sim$asusco/research/uband/red1/}}\\[-8pt]

\noindent and at:

{\footnotesize\sf
\indent\url{presto.stsci.edu/apsb/doc/pep/public-proposals/8645.prop}}


\section{\bfseries\sc\large Presentation of the WFPC2 Mid-UV Images}
\label{WFPC2MidUV-OpticalATLAS} 

In this section, we describe the \HST/WFPC2 images and discuss each
individual object in the sample.  In view of the large volume of the
combined ground- plus space-based data set, we opted to only present the
Cycle~9 \HST/WFPC2 images in the printed version of the present paper
(Figs.~3).  The electronic version contains the full ``pan-chromatic
atlas of nearby galaxies'', that include \UIT\ and ground-based
$UBVR(I)$ and $JHK$ images, wherever available.  Of these, we show in
the printed version a few example pages (Figs.~4) to clarify the adopted
format. 

The galaxies are presented roughly in order of their observed
morphological type.  We start with the regular early-type galaxies in
\S~3.01--3.03 (and correspondingly numbered figures Figs.~3.01--3.03 and
4.01--4.03).  Subsequent sections present early-type galaxies with
significant AGN in the mid-UV (\S~3.04--3.06), and merger products which
are likely to evolve into early-type remnants (\S~3.07--3.09).  Next, we
discuss early- through mid-type spiral galaxies showing spiral structure
in the mid-UV (\S~3.10), grand design spirals (\S~3.11--3.14), a spiral
with an inner ring (\S~3.15), and edge-on spiral galaxies
(\S~3.16--3.20).  We discuss the late-type- and dwarf spiral galaxies in
\S~3.21--3.27, and the Magellanic Irregulars in \S~3.27--3.33.  We
finish with the peculiar galaxies and merging/interacting systems
(\S~3.34--3.38). 

Although they are not part of the selected sample, other galaxies that
appear within the WFPC2 FOV will be discussed below (including their
types), in context with the targeted galaxies of similar morphological
type.  In a few cases, our \HST\ and ground-based images show the type
listed in the RC3 to be incorrect, for various reasons explained below. 

Most of our Cycle~9 mid-UV galaxies, except for some of their outskirts,
fit inside a single WFPC2 CCD, and so the current paper shows only a
single WFPC2 CCD in most Figures 3.nn--4.nn.  However, a few edge-on
galaxies are too large to fit on a single WFPC2 CCD.  For these
galaxies, the relevant parts of a full WFPC2 CCD \emph{mosaic} are
therefore shown instead in the current paper (see UGC~10043 in
Fig.~3.16).  Since the number of pixels we can display in this manner on
one ApJ page does not do full justice to the high quality of the WFPC2
data, we will display in a sequel paper (Odewahn \etal 2002b) the
multi-color WFPC2 mosaics in full-page size for the two remaining large
edge-on galaxies (\ie, ESO~033$-$G022 in Fig.~3.19 and IC~4394 in
Fig.~3.20) and possibly for a few other large objects, as needed.  This
sequel paper will dither all WFPC2 mosaics for our sample of 37 objects,
and do a detailed sky-subtraction across the CCD's, which is needed for
reliable determination of the mosaic-wide multi-color light-profiles, as
well as the subsequent quantitative analysis.  This paper will thus show
a few additional images not shown in the current paper. 

The \HST\ images, in general, have a very high dynamic range.  This
makes it difficult to properly display the full range of structures
within a galaxy.  Each of the grey-scale renditions in the atlas is
displayed using an inverse logarithmic stretch with lower and upper
display limits approximately equal to the sky-background
(\emph{displayed as white}) and the 3 times the maximum data value in
the galaxies' brightest parts (\emph{black}).  This compromise allows
one to discern individual high-SB structures within a galaxy while still
being able to see much of the faint outer parts within a galaxy.  The
two unexposed edges of each WFPC2 image allow one to trace the total
$80\arcsec\times80\arcsec$ ($75\arcsec\times75\arcsec$ useful) FOV of a
WFC CCD.  When a galaxy is much smaller than the WFC FOV, or when there
is too much morphological information contained within a single WFPC2
CCD, we (also) present the relevant portion of that CCD-frame, but
enlarged by a factor of 2 or 4 (\ie, a $37\arcsec\times37\arcsec$ or
$19\arcsec\times19\arcsec$ FOV). 

The ground-based images (in Figs.~4) tend to have a much smaller dynamic
range (because of the much higher sky-background and lower resolution). 
Nonetheless, we use a similarly defined stretch and display limits. 
Often, the extent of a galaxy becomes more obvious in these ground-based
images than in the mid-UV images.  In Figs.~4.01--4.38 all images are
displayed on the same angular scale and at the same orientation, with
North up and East to the left. 

\label{NotesonIndividualObjects}

\mn{\sc Mid-UV--faint Early-type Galaxies:}

\sn\textbf{3.01\ \ CGCG~097$-$094}\ (T=$-5$; E)\ \ An early-type galaxy
in Abell~1367 that is faint in the mid-UV compared to F814W. 
CGCG~097$-$094 was observed in the field of primary target
MCG+03$-$30$-$071.  Other cluster members are also visible (Figs.~3.01
and 4.01): the edge-on galaxy is still visible in F300W, suggesting
that, despite its dust content, it still transmits some light at
2930\AA.  Most of this UV light may come just from the near edge of the
disk (the side facing us).  MCG+03$-$30$-$071 itself is discussed in
\S~3.21, and edge-on galaxies are discussed further in \S~3.16--3.20
below. 

The vertical image flaws (faint streaks) in the F300W image are due to
WFPC2 Charge Transfer Efficiency (CTE) effects, the severity of which
has significantly increased over the past years.  They are visible in
the high contrast stretches that were needed to show the fainter
features of this and other objects in the atlas. 

\sn\textbf{3.02\ \ NGC~1396}\ (T=$-3$; S0$-$)\ \ A lenticular galaxy
that is faint in the mid-UV compared to F814W, indicative of an old
stellar population.  It has a weak nuclear point source.  NGC~1396 is
not detected at 150~nm in the \UIT\ far-UV image of its bright neighbor
NGC~1399, and is at best barely visible in the 250~nm \UIT\ image.  We
do not have ground-based images as yet for this galaxy, and so
substitute SERC IIIaJ and POSS-II IIIaF plate scans for the $B$ and $R$
images, respectively. 

\sn\textbf{3.03\ \ NGC~4478}\ (T=$-5$; E)\ \ An elliptical galaxy that
is faint in the mid-UV compared to F814W, although comparatively not as
faint in F300W as some of the other early-type galaxies discussed. 
NGC~4478 is one of the giant ellipticals in the Virgo cluster, many of
which have a UV-upturn in their far-UV spectrum (Burstein \etal 1988). 
This UV-upturn population is expected to be mostly visible below
2000\AA\ and contributes less than 3--30\%, and in most cases less than
10\% of the total light at 2930\AA\ (see Figs.~6--7 of Burstein \etal
1988).  Another point source is seen just below the nucleus.  It is very
bright at F300W and in the \UIT\ 250~nm image, and is likely a blue
foreground star. 

\mn{\sc Early-type Galaxies with Significant AGN Emission in the Mid-UV:}

\sn\textbf{3.04\ \ NGC~3516}\ (T=$-2$; S0)\ \ NGC~3516 is a well known
classical Seyfert~1 galaxy (Seyfert 1943; Khachikian 1974; Keel \&
Weedman 1978; Kent 1985; Filippenko, 1985) that becomes essentially a
point source in the \HST\ mid-UV images.  The AGN-dominated F300W image
shows a strong central point source exhibiting diffraction spikes.  A
faint extension is seen in this filter, but not in F255W.  The faint
fuzz seen around the central point-source in the 150 nm UIT image is
likely due in a significant part to the PSF caused by the far-UV optics
plus photographic film of the \UIT\ camera.  The ground-based and F814W
images show a faint bar with a position angle PA\,$\simeq\!-17$\arcdeg. 
NGC~3516 has a companion galaxy with a small velocity difference (Keel
1996).  It is also a well-known radio- and X-ray source (\eg, Taylor
\etal 1996; White \& Becker 1992; Laurent-Muehleisen 1997; Perlman \etal
1998; Radecke 1997), and is a low-luminosity AGN with the jet pointing
close to the line-of-sight to the observer.  Quantitative PSF-fits of
all these AGN dominated objects will be given by Odewahn \etal (2002, in
prep.), in conjunction with the \HST+ground-based light-profiles for all
our Cycle~9 sample. 

\sn\textbf{3.05\ \ UGC~03426}\ (\aka Mrk~3; T=$-2$; S0)\ \ An early-type
Seyfert~2 galaxy with a blue, nuclear ``disk-like'' feature that may be
associated with a bipolar outflow (Ruiz \etal 2001) rather than a
nuclear star-forming disk.  The feature is oriented roughly
perpendicular to the galaxy major axis in F814W and is visible in both
F300W and F255W.  Note the apparent change in galaxy position angle
blueward of the B filter in Fig.~4.05. 

\sn\textbf{3.06\ \ UGC~08823}\ (\aka Mrk~279; T=$-2$; S0)\ \ An
early-type galaxy that becomes essentially a point source in the mid-UV. 
Note that almost no extended light is seen in F300W compared to F814W. 
It appears to be interacting with its neighbor MCG+12$-$13$-$024, with
which it forms a physical pair (Keel 1996).  It is a well known
Seyfert~1 galaxy (Khachikian 1974; Osterbrock, 1977).  For a
high-resolution optical study of this object we refer the reader to
Knapen, Shlosman, \& Peletier (2000). 

\mn{\sc Early-type Galaxies with Evidence for Recent Merging or Strong 
Interaction:}

\sn The next three systems (sections 3.07--3.09) include morphologically
peculiar objects with single stellar systems.  We discuss these objects
here, at the end of the sequence of early-types, because they will
likely evolve into early-types on relatively short time scales (\lsim
1~Gyr; see Barnes \& Hernquist 1992), and likely already have developed
$r^{1/4}$ profiles in their inner parts (\eg, van~Albada 1982, Windhorst
\etal 1994b, 1998).  Earlier stage interactions --- in which the two
galaxies are still separated --- are discussed at the end of the
sequence (sections 3.33--4.38), after the Spirals and Irregulars, since
at high redshift such objects are more likely seen as interacting
galaxies or peculiars. 

\sn\textbf{3.07\ \ NGC~3921}\ (\aka Arp~224, Mrk~430, UGC~08823; T=0;
S0a)\ \ NGC~3921 is a disturbed early-type galaxy with a blue nuclear
region which is visible in F300W.  This object is a well-studied merger
remnant (Schweizer 1996, Yun \& Hibbard 2001a), and was probably
produced by the merger of a gas-rich and gas-poor progenitor (Hibbard \&
van Gorkom 1996, Schweizer \etal 1996).  It lies near the end of the
Toomre Sequence of on-going mergers (Toomre 1977), and represents the
final stage of merging, where there is a single nucleus but multiple
tidal features.  It will likely evolve into an elliptical galaxy
(Schweizer 1996, Hibbard \& Yun 1999a).  The ground-based images in
Fig.~4.07 show the fainter debris beyond the nuclear regions, including
remaining tidal features.  The near-IR images show a ``wake'' to the
North of the bulge --- possibly also a tidal feature from the same
merger. 

\sn\textbf{3.08\ \ UGC~05101}\ (T=11; Pec)\ \ This object is very faint
in F300W and is not detected in F255W.  The F814W images show that it
has an inclined dusty disk.  This object is an Ultra-Luminous Infrared
Galaxy (ULIG) with $L_{\mbox{\scriptsize IR}} \gsim 10^{12}\, L_{\odot}$
(Sanders \etal 1988).  Like NGC~3921, it appears to be a late stage
merger remnant, with a single nucleus (Scoville \etal 2000) and multiple
extended tidal features seen at larger scales in the ground-based
images, including a tidal tail and a loop or polar ring-like structure
(Sanders \etal 1988; Surace, Sanders \& Evans 2000).  The latter authors
explain the large scale morphology as resulting from a ``plunging''
collision at non-zero impact parameter of two highly inclined disks. 

Structures like these are typical for a merger in progress.  The tails
are thought to evolve on time-scales of $>$1~Gyr while, once the halos
overlap, the galaxy nuclei evolve on much shorter time-scales (of order
100~Myr).  This results in one visible tidal tail per progenitor disk,
while the other side of the disk gets pulled into a ``bridge'' before it
disperses.  A dusty disk, such as seen in UGC~05101, is a common feature
of gas-rich merger remnants, \eg, NGC~2623 (Bryant \& Scoville 1999),
NGC~3256 (Zepf \etal 1999), NGC~7252 (Wang, Schweizer \& Scoville 1992;
Whitmore \etal~1993), and Mrk~273 and NGC~3310 (this paper).  Disks have
recently been reproduced in numerical simulations of merging galaxies
(Naab \& Burkert 2001; Barnes 2002), and hence their presence does not
rule out a major merger origin, as was once thought. 

\sn\textbf{3.09\ \ UGC~08696}\ (\aka Mrk~273, VV~851; T=11; Pec)\ \ This
object is faint in F300W and barely detected in F255W.  The F300W
morphology is very similar to the $U'$ morphology presented in Surace \&
Sanders (2000).  It appears to have a dusty disk, visible on one side of
the nucleus in F814W.  It is a merger product, as attested to by the
tidal tail evident in F814W and in the ground-based images (Fig.~4.09). 
The horizontal streak in the ground-based R-band image is due to a
bright star outside the FOV.  Like UGC~05101, this system is a well
known and well studied ULIG.  But unlike the previous two systems, radio
and near-IR imaging reveal two distinct nuclei separated by 1$''$
(Majewski \etal 1993; Knapen \etal 1997; Scoville \etal 2000).  One of
the near-IR and radio nuclei coincides with the brightest peak in the
F300W image, but the second near-IR nucleus is obscured, even at F814W. 
The southern peak in the F300W image has no optical or near-IR
counterpart, and is probably a very young star cluster (Surace \&
Sanders 2000).  CO-line mapping reveals the presence of two orthogonal
kinematic components (Yun \& Scoville 1995; Downes \& Solomon 1998),
suggesting a highly inclined encounter between two gas-rich system. 

\mn{\sc Early-type Galaxies Displaying Spiral Structure in the Mid-UV:}

\sn\textbf{3.10\ \ NGC~2551}\ (T=0.2; S0a)\ \ An early-type spiral
galaxy, whose spiral structure is much more pronounced in F300W than in
F814W.  In the optical/red, the galaxy would be classified as an S0a, in
the mid-UV as an Scd, so its morphological K-correction is significant. 
NGC~2551's bulge is faint in the mid-UV compared to F814W, although it
seems bluer than that of other early-type spirals.  The background
galaxy seen in F814W does not appear in F300W, but this does not
necessarily mean that the disk of NGC~2551 is very dusty at this
location --- the background galaxy may just be as red as some of the
other field galaxies seen in the background of our other images (we
chose the exposure times in F300W and F814W to provide relatively high
S/N for nearby galaxies, not for higher redshift ones).  Fig.~4.10 shows
that NGC~2551 was also detected with \UIT\ at 250 nm, but probably not
at 150 nm.  The \UIT\ resolution (FWHM $\sim$3\arcsec) is much lower
than \HST's, but on large scales, the \UIT\ 250 nm morphology is
comparable to our \HST\ F300W image. 

\mn{\sc Grand-design Spiral Galaxies:}

\sn\textbf{3.11\ \ NGC~6753}\ (T=3; Sb)\ \ Grand-design spiral galaxy
with very pronounced spiral structure in both F814W and F300W. 
Dust-lanes are visible crossing the spiral arms in F814W, corresponding
to regions of reduced or no light in F300W.  A detailed discussion of
the \HST\ images of this object is given by Eskridge \etal (2002a).  An
enlargement of Fig.~3.11\emph{a} is shown in Fig.~3.11\emph{b}.  In
addition to dust lanes crossing the spiral arms in a regular pattern and
with a fairly constant pitch angle, there is a large dust-lane crossing
the inner spiral arms under a significantly different pitch angle, best
seen in the F814W image.  This is likely a foreground trail of dust (see
also \S~3.38).  Fig.~4.11 shows that the morphological K-correction is
modest from F300W through the $I$-band, although in the near-IR the
object appears to be of earlier type in ground-based seeing. 

\sn\textbf{3.12\ \ NGC~7769}\ (\aka Mrk~9005; T=3; Sb)\ \ Grand-design
spiral galaxy with very significant spiral structure in F300W, which is
more pronounced than that seen in F814W, suggesting active
star-formation.  In the optical/red, the galaxy would be classified as
an Sb, in the mid-UV as an Scd, so its morphological K-correction is
significant.  The enlargement in Fig.~3.12\emph{b} shows dust-lanes
crossing the spiral arms in F814W, corresponding to regions of reduced
or no light in F300W.  The object also shows a faint nuclear dust-lane
crossing its small bulge.  NGC~7769 is interacting with SBa galaxy
NGC~7771.  \HI\ mapping observations by Nordgren \etal (1997) show
NGC~7771 to have an extended gaseous tail, while NGC~7769
counter-rotates with respect to NGC~7771.  Therefore, this is thought to
be a prograde-retrograde encounter, with NGC~7769 having retrograde
kinematics.  Note the possible emergence of a bar in the ground-based
near-IR $J$ and $H$-band images. 

\sn\textbf{3.13\ \ NGC~3310}\ (\aka Arp~217, VV~356; T=4; Sbc)\ \ This
mid-type spiral galaxy is well-known to harbor a substantial global
star-burst, likely either the remnant of a merger or a significant
accretion event (Balick \& Heckman 1981; Mulder, van Driel, \& Braine
1995; Smith \etal 1996b; Kregel \& Sancisi 2001; and references
therein).  Intense star formation is readily visible in the nuclear ring
(Conselice \etal 2000), the spiral arms, and linear ``arrow'' feature
(\eg, Mulder \& van Driel 1996).  \HI\ mapping of this object shows
extended \HI\ tails (Kregel \& Sancisi 2001), and it is likely that the
spiral ``arms'' do not lie in a single plane.  As a result of the global
star-burst, the morphology of NGC~3310 remains similar in the F300W and
F814W images.  Some differences in structure between the red and mid-UV
result from dust features, however.  The enlargement in
Fig.~3.13\emph{b} shows that the galaxy nucleus appears displaced from
the line of symmetry of the inner spiral arms.  Fig.~3.13\emph{b} also
shows a small nuclear dust-lane crossing its small bulge, and dust-lanes
crossing the spiral arms (some at significant pitch-angles) in F814W,
corresponding to regions of reduced or no light in F300W. 

The ground-based images in Fig.~4.13 show that the outer spiral arms are
more complicated, and appear to dissolve in the southern part at all
wavelengths.  The far-UV morphology of NGC~3310 at \UIT\ spatial
resolution (Smith \etal 1996b) is also shown for comparison.  This is an
example of a galaxy that looks rather similar in all filters from the
far-UV through the red. 

\sn\textbf{3.14\ \ NGC~7685}\ (T=5.3; Sc)\ \ Barred late-type spiral
with significant spiral structure in F300W and F814W.  A small bar and
nuclear bulge are visible in F814W.  The bulge appears to be bisected by
a small nuclear dust lane.  Fig.~4.14 shows that the morphological
K-correction is modest from F300W through the $K$-band. 

\mn{\sc Spiral galaxies with inner rings:}

\sn\textbf{3.15\ \ NGC~6782}\ (T=0.8; Sa)\ \ Early-type spiral galaxy
with a spectacular ring structure in the mid-UV, visible in both F300W
and F255W.  An enlargement of Fig.~3.15\emph{a} is shown in
Fig.~3.15\emph{b}.  This ring is surrounded by two faint and apparently
partially dusty spiral arms as visible in F814W, and an outer spiral
structure and outer ring that is visible in F300W and in the optical
pass-bands (Fig.~4.15).  The inner ring appears to be driven by a small
bar visible in F814W and F300W at PA\,$=\!-30$\degree.  The ground-based
images in Fig.~4.15 show that the outer ring encloses an outer bar, with
a position angle (PA\,$\simeq\!0$\degree) that differs from that of the
inner bar visible in the \HST\ images.  A detailed discussion of the
\HST\ images of this object is given by Eskridge \etal (2002a).  In
depth studies of galaxies with such star-forming rings are given by Buta
\etal (\eg, 1996, 1998ab, 2000, 2001).  Fig.~4.15 suggests a significant
morphological K-correction from F255W through the $K$-band, most of
which is caused by the ring, but with the bars being more prevalent at
the longer wavelengths, as noticed in the ground-based near-IR study of
Eskridge \etal (2000). 

\mn{\sc Edge-on spiral galaxies:}

Here, we discuss edge-on galaxies as a single morphological class to
allow easier comparison, despite the fact that their intrinsic T-types
may span most of the Hubble sequence. 

\sn\textbf{3.16\ \ UGC~10043}\ (T=4.0; Sbc pec)\ \ An edge-on spiral
galaxy that is faint in F300W, although it is still visible in this
filter almost throughout the entire disk.  The central bulge is
relatively small in size and faint in the mid-UV, as for most of the
other edge-on galaxies in our sample.  The PA of the (major axis of the)
bulge is approximately perpendicular to that of the disk, which is most
clearly seen in the \HST\ observations when compared to the ground-based
images (Fig.~4.16).  This may indicate a polar ring, which
interpretation we hope to confirm with spectroscopy in a future paper. 
The F814W images show significant dust lanes, mostly in the inner parts
of the edge-on disk.  Compared to F814W, the F300W flux shines through
better in the outskirts than in the central part of the disk, and in
particular, it shines through well in several bright knots.  It is also
possible that, instead, the knots visible in the mid-UV are in front of
most of the dust. 

Qualitatively, it appears that the dust extinction --- if this relative
dimming of the UV-light is indeed due to dust extinction --- decreases
from the inside outwards (see also Jansen \etal 1994).  Examples of
these edge-on objects have been studied in the optical by de Grijs
(1999).  In the mid-UV, edge-on disk galaxies have a large range in
brightness.  Some are particularly bright in the UV (\eg, UGC~06697,
Marcum \etal 2001; NGC~4631, Smith \etal 2001), while others are much
fainter, such as some of the edge-on galaxies shown in the current
paper.  As pointed out by Kuchinski \etal (2001), it is as yet unclear
to which degree dust attenuation affects the appearance of highly
inclined galaxies at UV wavelengths.  This is due to a likely complex
dependence on the distribution of dust and actively star-forming regions
throughout a galactic disk (Kuchinski \etal 2001).  Due to the lack of a
statistically significant population of edge-on galaxies observed in
mid-UV light, a detailed discussion of this effect is beyond the scope
of the present paper, but will be addressed in a future paper when more
edge-on objects have been observed in the mid-UV. 

\sn\textbf{3.17\ \ UGC~06697}\ (T=10; Im)\ \ Late-type edge-on galaxy
with almost no bulge and with significant dust-lanes seen by comparing
F300W to F814W.  Although this galaxy is bright in the UV and was
observed both at 2500\AA\ and 1500\AA\ by \UIT, comparison of the F300W
and F814W images shows that several regions seem to be completely
obscured by dust.  They appear to be dust pockets that bisect the disk. 
The ground-based images in Fig.~4.17 show a significant warp on larger
scales, which is possibly related to the two or three companion galaxies
visible within the FOV.  On balance, though, the overall appearance of
this object is very similar from the far-UV through F814W. 

\sn\textbf{3.18\ \ IC~3949}\ (T=$-2$; S0 pec)\ \ Edge-on spiral with a
very faint bulge, if any, and significant dust-lanes (compare the F300W
and F814W image), especially in the lower (western) part of the
displayed image.  The plane of the dust appears to be warped, and
asymmetric in the western part of the image compared to the upper
(eastern) part.  The biggest dust pocket to the lower right of the
center is also visible as a depression in the light in the ground-based
$UBVR$ images (Fig.~4.18). 

Four faint edge-on galaxies are seen to the left of IC~3949 in F814W
(Fig.~3.18\emph{a}), two of which are barely visible in F300W.  It is
perhaps noteworthy that all four of these highly inclined galaxies point
their major axis towards the central part of IC~3949.  Although this is
probably a chance alignment and the edge-on galaxies are probably
background galaxies, it would be interesting to test this hypothesis
with further kinematic data.  If these systems turn out to be at a
similar distance as IC~3949, they might be smaller dwarf-like objects
falling into IC~3949, perhaps disturbing its disk as the irregularly
distributed edge-on dust-lanes might suggest.  Schwarzkopf \& Dettmar
(2000) found that edge-on galaxies have on average 60\% thicker disks
and are $\sim 2\times$ more likely to have warps when they have low mass
companions, which will likely soon result in minor mergers. 

Fig.~3.18\emph{b} shows other early-type objects surrounding IC~3949,
which, like other early-type galaxies in our \HST\ sample, have dim
light distributions in F300W.  The RC3 incorrectly classifies IC~3949 as
S0 pec, which may be due to confusion with the brighter of these
early-type galaxies, although this object is fainter than the edge-on
galaxy.  These objects were selected as field ellipticals and do not
show AGN in the UV (unlike the others in our sample), although the two
brightest ones show a small central disk or boxy isophotal structure. 

\sn\textbf{3.19\ \ ESO~033$-$G022}\ (T=7; Sd)\ \ Edge-on late-type
spiral galaxy that is faint in F300W, although it is still barely
visible in this filter throughout most of the disk.  There is almost no
central bulge.  The F814W images show a thin dust lane in the plane of
the galaxy, mostly in the inner parts of the edge-on disk.  Compared to
F814W, the F300W flux shines through better in the outskirts than in the
central part of the disk, and in particular shines through well in
several bright knots.  In the ground-based images for this low Galactic
latitude object from de~Grijs, Peletier \& van~der~Kruit (1997) (see
Fig.~4.19), the bright stars (which are apparent in the F814W \HST\
image) have been removed to allow for more accurate surface photometry. 

\sn\textbf{3.20\ \ IC~4394}\ (\aka ESO~446$-$G044; T=6; Scd)\ \ Edge-on
spiral galaxy that shows a thick dust lane in the plane of the galaxy in
F814W, mostly in the inner parts of the edge-on disk.  This galaxy is
fainter in F300W than F814W as well, but comparatively not nearly as
much as most of the other edge-on galaxies.  This could be because its
inclination is not as close to 90\arcdeg as some of the others, allowing
a more unimpeded view of the bright star forming regions located on the
near side of the galaxy plane.  Hence, the relative bright appearance in
F300W does not imply a smaller dust content than inferred in other
edge-ons.  Almost no central bulge is visible.  Compared to F814W, the
F300W flux shines through better in the outskirts than in the central
parts of the disk, and in particular in several bright knots, more so
than seen in the other galaxies. 

\mn{\sc Late-type spirals and smaller spirals:}

\sn\textbf{3.21\ \ MCG+03$-$30$-$071}\ (T=10; Im)\ \ Small late-type
spiral galaxy in Abell~1367, displaying a four-armed spiral pattern that
dissolves into individual star forming knots in F300W. 
MCG+03$-$30$-$071 was detected by \UIT\ both at 1500\AA\ and at 2500\AA. 
There are several galaxies at small projected distances to this galaxy,
but the smallest difference in velocity is 715 km s$^{-1}$ (for the
early-type neighbor \tsim40\arcsec\ to the SW, MCG+03$-$30$-$067, which
is also visible in the FOV).  Comparing these two galaxies from the
far--UV to F814W gives the best example of the general trends observed
in this work (see the discussion in \S~\ref{DiscussionofGeneralTrends}). 

\sn\textbf{3.22\ \ MCG+06$-$24$-$047}\ (T=4; SBc)\ \ Small spiral galaxy
with well-developed spiral structure that is more pronounced in F300W,
although still clearly visible in F814W.  The classification in the RC3
as T=11/Pec probably resulted from the bright foreground star \tsim
7\arcsec\ south of the nucleus.  In the mid-UV this galaxy would be
classified as a Sd/Irr rather than SBc, so the morphological
K-correction is non-zero.  The ground-based $UBVR$ images (Fig.~4.22)
show a rather modest dependence of the amplitude of the spiral structure
on rest-frame wavelength, so most of the morphological K-correction is
caused by the stellar population below the Balmer break.  A small bulge
is visible in F814W.  The bright star is blue (it is even marginally
detected at 1500\AA), so that the F300W appearance is not significantly
affected by the filters' red-leak. 

\sn\textbf{3.23\ \ UGC~05028/29}\ (\aka Arp~300, VV~106; T=3+9;
Sbc+SBdm/Pec)\ \ Fig.~3.23\emph{a} shows the small late-type/peculiar
spiral galaxy UGC~05028 (Mrk~111) of the pair Arp~300 (see Fig.~4.23). 
It has spiral structure that is more pronounced in F300W although still
clearly visible in F814W.  Ground-based seeing hides most of the
irregular nature of this object.  No bulge is visible in F814W, although
a small central bar-like structure is visible in both filters.  The
brightest ``knot'' in F300W and particularly in F814W, south of the
center, is well resolved.  Although reddening by dust plays a
significant role, this ``knot'' may be the remnant of another small
late-type galaxy that is merging with UGC~05028, as its color is similar
to that of the small bulge of UGC~05028's larger companion UGC~05029. 
If so, this knot and the bar will eventually merge and form a central
bulge.  As presently observed UGC~05028 is extremely asymmetric. 

Fig.~3.23\emph{b} shows the second member of the Arp~300 system, Sbc
spiral galaxy UGC~05029, which is a physical companion to UGC~05028. 
The spiral structure in this galaxy is more pronounced in F300W,
although it is clearly also visible in F814W, like for some of the other
early--mid type spirals discussed above (\eg, NGC~2551).  Most of the
blue OB associations are located on the side facing UGC~05028.  Another
edge-on field galaxy is seen in Fig.~4.23, but is too faint to be
resolved into star-forming regions.  The five rather red objects between
this edge-on galaxy and the bright mid-type spiral are likely a group of
background galaxies --- they are not seen in F300W at all (see the
discussion of other background objects above). 

\sn\textbf{3.24\ \ NGC~3860B}\ (\aka CGCG~97$-$114; T=10; Im)\ \ Small
late-type spiral galaxy with spiral structure that is more pronounced in
F300W, although it is still visible in F814W.  Its rather late
classification probably resulted from the fact that the spiral structure
is only barely visible in the ground-based images.  The \UIT, \HST\ and
ground-based images show only a modest dependence of morphology on
rest-frame wavelength.  A very small bulge is visible in F814W
(Fig.~3.24). 

\sn\textbf{3.25\ \ ESO~418$-$G008}\ (T=8; Sdm)\ \ Small late-type spiral
galaxy that is resolved in at least its brighter star-forming regions
(Fig.~3.25), which are mostly distributed along its outer perimeter, and
perhaps constitute the beginning of spiral structure.  The F814W image
also suggests a faint bar running along the minor axis that may be
connected to this spiral structure.  The $B$ and $V$ images of Matthews
\& Gallagher (1997) show that ESO~418$-$G008 has a bright, prominent bar
(see also Fig.~4.25).  A faint low-SB structure is visible in F814W that
is spread more smoothly throughout the galaxy, presumably its unresolved
older stellar population.  A color map made from our \HST\ data
({\sf\url{oposite.stsci.edu/pubinfo/pr/2001/04/}}) reveals a very faint
red nucleus.  Such compact nuclei are common features of extreme
late-type spirals (Matthews \& Gallagher 1997; Matthews \etal 1999;
B\"oker \etal 2002).  The nucleus of ESO~418-G008 was not previously
visible in ground-based images (Matthews \& Gallagher 1997), but B\"oker
\etal (2002) have recently imaged it in the $F814W$ filter with the
WFPC2 PC.  They measure $M_I=-10.24$ mag, assuming a distance of 14.1
Mpc. 

The galaxy appears to have sharp ``edges'' in both filters, but this may
be largely a visual impression caused by the single star-forming region
to the lower left (north) in Fig.~3.25.  The ground-based $UBVR$ images
show similarly an outer spiral-like structure that becomes less and an
inner bar that becomes more pronounced at longer wavelengths.  A
late-type galaxy with a modest morphological K-correction. 

\sn\textbf{3.26\ \ NGC~1510}\ (T=-2.3; S0 pec)\ \ Small blue, amorphous
galaxy (Eichendorf \& Nieto 1984).  NGC~1510 is a physical companion to
the much larger SB galaxy NGC~1512, with which it is interacting
(Hawarden \etal 1979).  The early-type classification in the RC3
probably resulted from ground-based seeing hiding most of the irregular
nature of this object.  No clear bulge is visible.  NGC~1510 has fairly
similar appearance in F814W and F300W, but the F814W image shows more of
a diffuse component, while the F300W resembles more a disk-like
component seen edge-on, and shows some linear dust features.  The object
is resolved in at least its brighter star-forming regions.  Scans of
ground-based photographic plates in $B$ (from Lauberts \& Valentijn
1989, as obtained through NED) and in $R_F$ (from the second generation
Digital Sky Survey) are substituted for CCD images in Fig.~4.26. 

\mn{\sc Magellanic Irregulars:}

\sn\textbf{3.27\ \ NGC~5253}\ (\aka Haro 10; T=10; Im)\ \ NGC~5253 is a
peculiar I0 galaxy or type II irregular.  This object has a very similar
appearance in F300W and F814W, and is clearly resolved into individual
bright star-forming regions or stars.  Table~1 shows that it is one of
the closest galaxies in our sample.  The dust lane that appears to
emerge from one side of the center outwards is visible at all
wavelengths, including in the lower-resolution ground-based $UBVR$
images (Fig.~4.27).  Overall, this object has a largely similar
appearance from \UIT\ 150 nm through the ground-based $K$-band, with
some mild wavelength dependence modulations due to dust patches. 

This object was first described after discovery of a nearby nova (Hubble
\& Lundmark, 1923; Payne-Gaposchkin, 1936).  NGC~5253 is a a well
studied nearby star-bursting galaxy at a distance of 4.1~Mpc (Sandage
\etal 1994).  It is typically classified as an amorphous or irregular
galaxy, but has outer isophotes reminiscent of a dwarf elliptical
(Sersic \etal 1972; Caldwell \& Phillips 1989).  It has a star-burst
located at its center, with dozens of intense blue star forming clusters
(Caldwell \& Phillips 1989; Storchi-Bergmann \etal 1995; Calzetti \etal
1995).  NGC~5253 also contains an unusual H$\alpha$ morphology (Marlowe
\etal 1995; Calzetti \etal 1995) with loops and filaments and a diffuse
component that is at least partially produced from shocks (Calzetti
\etal 1999). 

Recent \HST\ images in optical broad-band and line-emission show that
this object is a small star-bursting galaxy (Beck \etal 1996; Calzetti
\etal 1997, 1999; Gorjian, 1996).  The galaxy was detected by IRAS
(IRAS~13370-3123 in the IRAS Point Source Catalog; Beichman \etal 1988),
and is a weak radio source in the Parkes-MIT-NRAO 4.85~GHz Survey
(Wright \etal 1996), and also had a FAUST Far-UV point source
0.5\arcmin\ from the galaxy (Bowyer \etal 1994).  Other recent studies
include a systematic \HI\ study (Kobulnicky \& Skillman 1995), radio
observations of a star-bursting knot that has no optical counterpart
(Turner \etal 1998, 2000), as well as a CO-map (Turner \etal 1997),
ROSAT X-ray studies of multiple super-bubbles in its star-burst nucleus
(Strickland 1999), and ISO observations of its hot stars and Wolf-Rayet
like outflow (Crowther \etal 1999), amongst others. 

Only 130~kpc away is the larger spiral galaxy NGC~5236 (M~83) which also
experiences a star-burst in its nuclear regions.  NGC~5253 has a large
outer distribution of \HI\ gas that could be feeding the central
star-burst in M~83 (Kobulnicky \& Skillman 1995).  Also, the bar in M~83
might be triggering its central star-burst.  However, there is evidence
that the star-bursting properties in this system are produced in part by
a dynamical interaction between the two objects (Caldwell \& Phillips
1989). 

\sn\textbf{3.28\ \ NGC~1140}\ (\aka Mrk~1063, VV~482; T=10; Im)\ \
Irregular galaxy.  This object has a fairly similar appearance from the
mid-UV through the $I$-band (Fig.~4.28).  In F300W and F814W (Fig.~3.28)
the object is resolved in at least its brighter star-forming regions. 
The ``super star clusters'' visible here were studied using \HST\
optical-band imaging by Hunter, O'Connell, \& Gallagher (1994).  The
larger field-of-view in the ground-based images also shows a trail of
knots to the south-west, possibly an infalling dwarf companion. 

\sn\textbf{3.29\ \ UGC~05626}\ (T=10; Im)\ \ Irregular galaxy.  The
object is resolved in at least its brighter star-forming regions which
are mostly draped around its edges as seen in F300W.  The bluest and
brightest star-forming knots to the left (north) are also visible in
F255W.  In F814W, the object also shows a faint low-SB structure which
is spread more smoothly throughout the galaxy, presumably its unresolved
older stellar population.  No central bulge is visible at any
wavelength.  The structure in the ground-based $UBVR$ images (Fig.~4.29)
is similarly independent of wavelength, and shows the outer rings of
knots, resembling the beginning of spiral structure.  It is curious that
such spiral structure would start before even a small central bulge or
bar has formed, and leads one to wonder about the dark matter content of
(the central parts of) this galaxy. 

\sn\textbf{3.30\ \ UGC~09855}\ (T=10; Im)\ \ Magellanic type irregular. 
The object is resolved in at least its brighter star-forming regions. 
The brightest of these regions are also visible in F255W.  The F814W
shows a faint central bulge that is not quite aligned with the
geometrical center of the object (see also the ground-based images in
Fig.~4.30, and the disk of UGC~05626 in \S~3.29). 

\sn\textbf{3.31\ \ NGC~6789}\ (T=10; Im)\ \ Magellanic type irregular. 
The object is resolved in at least its brighter star-forming regions. 
The brightest ones of these are also (but barely so) visible in F255W. 
The structure in the ground-based $UBVR$ images (Fig.~4.31) is rather
independent of wavelength, and no central bulge is visible at any
wavelength. 

\sn\textbf{3.32\ \ Mrk~66}\ (T=11; Pec)\ \ This object shows a
significant number of blue star-forming knots, the one to the left
(west) of the center (Fig.~3.31, F814W) may be part of a faint small
bulge.  The knots to the right (east) are isolated from the main object
in both filters, and may be separated from the main body by a dust-lane
or pocket, as is also seen in the ground-based images (Fig.~4.32).  It
is not clear whether this galaxy is a small linear object or a late-type
galaxy seen edge-on. 

The ground-based images provide some additional clues, since in all
$UBVR$ filters, a low-SB disk or halo is seen surrounding the rather
linear structure seen in F300W.  If this rounder feature were a halo,
then Mrk~66 could be a tri-axial amorphous galaxy, although the fact
that this feature is also visible in the ground-based $U$-band filter
below the 4000\AA-break would imply a rather young halo age of \lsim
1--2 Gyrs.  Alternatively, if this rounder feature is a face-on disk,
then the linear feature that is primarily visible in F300W could be a
star-forming bar.  In that case, the rounder feature would have to be a
rather old disk ($\sim$1 Gyr), since no spiral arm structure is visible
at the same location in the F300W image.  Spatially resolved
spectroscopy and kinematics will be necessary to distinguish between
these possibilities. 

In combination with the \HST\ F300W and F814W images, the ground-based
$UBVR$ images confirm that, to first order, this object has little
morphological K-correction, \ie, its appearance is rather similar at
most wavelengths.  However, significant SB-dimming at high redshifts
could change a roundish object like Mrk~66 as seen in visible light into
a more linear object in the mid-UV, and so help explain the appearance
of some of the ``chain'' galaxies seen at high redshifts (Cowie, Hu, \&
Songaila 1995; see also Dalcanton \& Shectman 1996). 

\sn\textbf{3.33\ \ UGC~05189}\ (\aka VV~547; T=10; Im)\ \ Like Mrk~66,
the \HST\ images of this object show a significant number of blue
star-forming knots.  The ground-based $UBVR$ images (Fig.~4.33) show
that the arc imaged with WFPC2 is in fact part of a much larger, low-SB
structure.  This is clearly a very dynamically disturbed system.  There
is no obvious central bulge.  In combination with the \HST\ F300W and
F814W images, the ground-based $UBVR$ images confirm that this object
has almost no morphological K-correction, \ie, its appearance is about
the same at all wavelengths.  Its light must be dominated by young hot
stars throughout. 

\mn{\sc Interacting, Peculiar, and Merging Galaxies:}

\sn\textbf{3.34\ \ NGC~3690/IC~0694}\ (\aka Arp~299, VV~118; T=9; Sm)\ \
Merger of two late-type disk galaxies.  The object has a very similar
morphology in F814W, F300W, and F255W, although the sensitivity in F255W
is much reduced.  Comparison of the F300W and F814W images reveals
significant dust patches, most of which are irregularly distributed, as
shown in the enlargements in Fig.~3.34\emph{b}.  This figure also shows
the numerous young super star clusters spread throughout the system
(\eg, Meurer \etal 1995). 

The objects somewhat resemble the Antennae Galaxies (NGC~4038/39), but
the encounter geometry is more complicated and the merger is slightly
more advanced in Arp~299.  \HI\ and CO kinematics suggest a
prograde-retrograde or prograde-polar encounter (Hibbard \& Yun 1999b,
Casoli \etal 1999).  There are no obvious bulges in the mid-UV--optical
images, but NIR imaging reveals a bulge in at least the easternmost
system (Smith \etal 1996a), which is quite hidden by dust in all filters
blueward of \tsim1\micron.  Several of the bluest star-forming knots
shine through clearly in F300W and F255W, although most of the galaxy
shows significantly dusty regions, where such star-bursting knots would
be much more reddened. 

\sn\textbf{3.35\ \ UGC~06527}\ (\aka Arp 322, Mrk~176, VV~150, HCG~056;
T=11; Pec)\ \ Spiral galaxy in a Hickson compact group that is
undergoing a merger with at least one of its companions (\eg, Fasano \&
Bettoni 1994, Allam \etal 1996).  Its spiral disk and arms are distorted
into tidal tails because of this encounter, as also implied by
UV+photographic numerical simulations of such events (\eg, Barnes \&
Hernquist 1992).  When comparing F814W to F300W, the inner bulge shows a
dusty inner disk-like structure.  The galaxy is not resolved into its
brightest star-forming regions or stars, due to its larger distance. 
The ground-based images in Fig.~4.35 suggests a small morphological
K-correction, if any. 

\sn\textbf{3.36\ \ UGC~08335}\ (\aka Arp~238, VV~250; T=4; Sbc)\ \ Two
strongly interacting disk galaxies of approximately equal luminosity. 
Both disks are distorted during this encounter and display tidal tails
(Fig.~4.36), suggesting a prograde-prograde encounter (\eg, Barnes \&
Hernquist 1992).  The RC3 morphological type probably results from
confusing the tidal arms with spiral structure or the two nuclei with a
central bar (Fig.~4.36) in ground-based images.  When comparing F814W to
F300W, the inner bulges of both systems show significant dust-lanes,
possibly from gas-rich material raised during the encounter.  The system
is too distant to be resolved into its brightest star-forming regions or
stars. 

\sn\textbf{3.37\ \ IC~2184}\ (\aka Mrk~8, VV~644; T=9; Sm)\ \ The
``flying V'' is probably a merger of two disk galaxies seen nearly
edge-on, as indicated by the presence of pronounced tidal tails in deep
ground-based images (Gallagher \etal 2000).  Star formation is active in
most of both of the disks, and in the F255W image only the brightest
star-forming knots are seen due to the reduced sensitivity of WFPC2 in
this filter.  Dust patches are visible in F814W, corresponding to
reduced or no light in F300W and F255W at those locations.  The
ground-based $UBVR$ images (Figs.~4.37) similarly show a very modest
dependence of the morphology on rest-frame wavelength. 

\sn\textbf{3.38\ \ NGC~5278/79}\ (\aka Arp~239, Mrk~271, VV~19; T=3;
Sb)\ \ This is one of the most curious objects in our sample.  Like
UGC~08335, this system consists of two strongly interacting spiral
galaxies.  The encounter clear distorts both galaxy disks, resulting in
tidal tails and exciting a strong $m=1$ mode in the larger of the two
galaxies, similar to features seen in numerical simulations of such
events (Zeltwanger, Comins, \& Lovelace, 2000; see also Phookun \etal
1992).  When comparing F814W to F300W, both the inner bulges show
significant dust-lanes.  Part of the dust appears to be pulled out along
with one of the spiral arms in the tidal encounter, possibly due to a
shock related to this encounter. 

Most remarkable is the very curved thin dust-lane, best seen in $R$ and
$F814W$, that drapes across the bigger galaxy.  This dust-lane starts
somewhere in the southern spiral arm of the bigger galaxy (upper or
southern part in the WFPC2 CCD as shown in Fig.~3.38), curves all the
way across the small nuclear bulge of this galaxy, and makes it into the
spiral arm to the lower left that appears to connect to the smaller
galaxy.  Careful comparison of the F300W and F814W images shows that the
dust-lane starts in what appears to be at least two funnel-like regions
from the southern arm of the bigger galaxy.  The dust in the southern
arm appears to be distributed non-uniformly along the arm in several
(3--4) sizeable ``dust pockets or patches'', which appear to merge into
the dust lane and tidal tail that wraps around the bigger galaxy's
nuclear bulge.  A somewhat similar but much thinner dust trail was seen
in \HST\ images of the colliding nearby galaxy pair NGC~1409/1420 (Keel
2000), with material being pulled out between the two galaxies. 

These ``dust pockets or patches'' are clearly visible in the data of the
F300W images, and at the same location, the F255W images are entirely
devoid of flux, leading credence to the conjecture that dust pockets
cause these depressions in the light distribution.  The ``patches'' are
best seen by blinking the F255W, F300W, and F814W images in rapid
succession, each displayed at an appropriate logarithmic stretch.  The
bluer ground-based images (Fig.~4.38) show a similarly diminished
surface brightness at the location of the dust patches.


\section{\bfseries\sc\large Discussion of General Trends in the WFPC2
Mid-UV Images}
\label{DiscussionofGeneralTrends}

Here we discuss the sub-samples of each of the Hubble types separately. 
The light-profiles of all types will be presented in a subsequent paper
in conjunction with the ground-based $UBVR$-profiles, after all the
ground-based zero-points have been secured (see Table~2).

\subsection{Early-type Galaxies}
\label{EarlytypeGalaxies}

The regular early-type galaxies (elliptical and S0's) in our sample show
in general a significant change in SB from the mid-UV to the red,
reflecting an overall lack of a young stellar population
(CGCG~097$-$094, NGC~1396, NGC~4478; sections 3.01--3.03).  Typically,
the \SB\ of E/S0's is 3--5 \magarc\ fainter than
$\mu_{\mbox{\scriptsize\sl F814W}}$.  As was also seen in the far-UV
\UIT\ sample of Marcum \etal (2001) and Kuchinski \etal (2000), they are
dominated by an older stellar population, and so are generally faint in
F300W.  Out of 7 early-type galaxies imaged, two have small blue nuclear
features (UGC~03426, NGC~3921; \S~3.05 and 3.07).  UGC~03426 is a
Seyfert~2, and so this feature could be associated with a bi-polar
outflow.  The other, NGC~3921, is a well studied merger-remnant from the
Toomre sequence, and the blue nuclear feature could indicate a location
of residual star-formation that is only partially obscured by dust. 
Three early-type galaxies (NGC~1396, NGC~3516, and UGC~08823; \S~3.02,
3.04, and 3.06) become dominated by point sources in the mid-UV,
indicating weak optical-UV AGN, Seyfert, or LINER nuclei.  While these
are small number statistics, their presence in our current small sample
of early-type galaxies is due to our selection of galaxies with high
predicted average SB in the mid-UV within their effective (\ie,
half-light) area (\S~\ref{HSTWFPC2SampleSelection}).  This resulted in
including a number of objects in our sample that are dominated in the
$U$-band and mid-UV by AGN. 

If AGN generally reside in bulge-dominated galaxies (\eg, Magorrian
\etal 1998), then the red and old stellar population of the underlying
early-type galaxies will be generally faint in the UV, but the presence
of a (weak) AGN will result in blue $(U-B)$ colors and therefore a high
SB the mid--near UV, and so inclusion into UV-selected samples. 
Similarly, ground-based U-band selected surveys would result in
significant numbers of AGN in unrecognized early-type galaxies at
moderate redshifts ($z\simeq 0.3$).  At redshifts of a few tenths, the
ground-based U-band selection would similarly show these objects
essentially as point sources, just like some of the nearby early-type
galaxies selected in our \HST\ mid-UV sample.  However, had these
objects been selected at $z\simeq 0.3$ from the ground at much longer
wavelengths (\ie, in the $I$-band), then they would still have shown up
as early-type galaxies in images with good ground-based seeing.  This
leads us to wonder to what extent the (strong) cosmological evolution of
AGN selected optically or through their near-UV excess at modest
redshifts (Koo \& Kron 1988; Boyle \& Terlevich 1998; Boyle \etal 2000)
could in part be due to a ``morphological K-correction'' of early-type
galaxies with weak AGN.  A further discussion of the galaxy light
underlying low redshift QSO's is given by Bahcall, Kirhakos \& Schneider
(1995, 1996, 1997), Kirhakos \etal (1999) and Maoz \etal (1996), who
determine the AGN components seen with WFPC2 in early---mid type
galaxies. 

Two more galaxies, UGC~05101 (\S~3.08) and UGC~08696 (\S~3.09), are
tentatively placed in this section of early-type galaxies.  These
objects are merger remnants and have significant dusty disks in the
mid-UV at present, but we believe they will soon evolve into early-type
galaxies.

\subsection{Mid-type Spirals}
\label{MidtypeSpirals}

In general, spiral arms are more pronounced in the mid-UV, as \UIT\ has
shown to be generally true in the far-UV (Bohlin \etal 1991, Hill \etal
1992, Kuchinski \etal 2001, Marcum \etal 2001), but mid-type spirals and
star-forming galaxies (\eg, NGC~6753 [\S~3.11] and NGC~3310 [\S~3.13])
appear more similar from the mid-UV to the optical than the early-type
galaxies discussed above.  However, several appear as later types in the
mid-UV (\eg, NGC~7769; \S~3.12), and a few show drastic changes in type
from the optical to the mid-UV, equivalent to an apparent change in
morphological type $\Delta T\gsim 3$ (\eg, NGC~2551 [\S~3.10] and
NGC~7685 [\S~3.14]).  One galaxy shows a spectacular resonance ring full
of hot stars (NGC~6782; \S~3.15), while the remainder of the disk
becomes essentially invisible in the mid-UV.  This object is discussed
in detail by Eskridge \etal (2002a). 

We see a variation in color of the galaxy bulges/centers in the spirals,
and a considerable range in scale and SB of the individual star-forming
forming regions.  Dust features in mid-type spirals can be well traced
by comparing the F300W to the F814W images.  Dust is visible in lanes or
patches (possibly trailing spiral density wave patterns?), in pockets,
and/or bubbles.  A curious feature is that almost without exception, the
mid-type spirals in our sample have their small nuclear bulges bisected
by a dust-lane, which is often connected to the inner spiral arm
structure.  This is interesting in the context of the recent finding
that all bulge-dominated systems have a central black-hole with an
average black hole mass $M_{bh}\simeq 0.005\times M_{bulge}$ (Magorrian
\etal 1998).  The small nuclear dust-lane seen by comparing the F814W
and mid-UV images of almost all mid--late-type spiral galaxies in our
sample may then be involved in feeding the inner accretion disk (which
is not visible in our \HST\ images). 

Some edge-on galaxies are very faint in F300W when compared to F814W
(UGC~10043 [\S~3.16] and ESO~033$-$G22 [\S~3.19]), others emit/transmit
significantly more in F300W (UGC~06697 [\S~3.17] and IC~3949 [\S~3.18]),
while IC~4393 (\S~3.20) is an intermediate case.  All show a F300W/F814W
ratio increasing from the inside out, as expected for a decreasing dust
content from the inside out, and/or a strong radial gradient in the
stellar population, if we assume that the dust and light sources are
well mixed (\ie, the light sources are not preferentially located in
front of most of the dust; see, \eg, Jansen \etal 1994; Witt \& Gordon
1996; Kuchinski \etal 1998).

\subsection{Late-type Galaxies and Irregulars} 
\label{LatetypeGalaxiesandIrregulars}

The late-type and irregular galaxies imaged are a heterogeneous mixture. 
The majority of these galaxies show a F300W morphology that is similar
to that in F814W.  Important differences are seen, however, due to
recognizable dust-lanes or pockets blocking out F300W light (\eg,
UGC~06697 [\S~3.17]; NGC~5253 [\S~3.27]; UGC~09855 [\S~3.30]; Mrk~66
[\S~3.32]; Arp~299 [\S~3.34]; IC~2184 [\S~3.37]).  Dust is visible in
pockets, holes or bubbles, perhaps due to supernova-induced outflows or
outflows fueled by bright star-forming regions, such as seen in M~82 at
\HST\ resolution (de Grijs \etal 2001). 

Some late-type galaxies are physically smaller galaxies with what
appears to be the beginning of spiral structure (MCG+03$-$30$-$071 and
MCG+06$-$24$-$047 [\S~3.21--3.22], UGC~05028 [\S~3.23], NGC~3860B
[\S~3.24], ESO~418$-$G008 [\S~3.25], UGC~05626 [\S~3.29]).  Others are
Magellanic Irregulars (NGC~5253, NGC~1140 and NGC~1510 [\S~3.26--3.28],
UGC~09855, NGC~6789, Mrk~66, and UGC~05189 [\S~3.30--3.33]) with various
regions of stochastic star-formation.  Star-formation ``ridges'' are
commonly seen in the late-type galaxies, as well as hot stars or
star-clusters that are particularly conspicuous in the mid-UV (F300W
and/or F255W). 

A few late-type galaxies would be classified significantly different
when observed in the mid-UV than in the F814W passband
(MCG+06$-$24$-$047 [\S~3.22], UGC~05626 [\S~3.29], UGC~09855 [\S~3.30]),
especially when observed under less than perfect atmospheric seeing
conditions.  A quantitative discussion of the classification changes as
a function of rest-frame wavelength will be given by Odewahn \etal
(2002b).

\subsection{Peculiars and Mergers}
\label{PeculiarsandMergers}

The majority of the peculiars and interacting galaxies (UGC~05189,
Arp~299, UGC~06527, UGC~08335, IC~2184, and Arp~239 [\S~3.33--3.38])
show a F300W morphology that is similar to that in F814W.  Several
mergers have spectacular dust-lanes that thread along with the spiral
arms, which are tidally distorted during the interaction (UGC~08335;
Arp~239), while most others have absorption in apparently random
locations associated with a star-burst. 

The \HST\ images are in general not sufficiently deep nor do they sample
a sufficiently large area on the sky to capture the lower SB features
(\eg, low-SB tails and other large tidal features, faint extended disks
or halos) that are well imaged on the ground-based photographic plates
which were originally used to classify these objects.  The inner regions
of these objects reveal a wealth of structure in the \HST\ images, but
often are almost entirely burned out or blurred on these ground-based
images.  The \HST\ images do thus not necessarily reflect the
distribution of the extended (evolved) stellar population.  This is
particularly true of the interacting systems, where tidal forces first
and most easily affect the outer regions of the participants.  Because
interacting galaxies have both a disturbed distribution of old stars in
their outskirts \emph{and} an irregular distribution of dust and young
star-forming regions in their inner parts, their peculiar morphology is
maintained over a range of spatial scales and surface brightnesses.  As
such, the morphological K-corrections derived from this sample mostly
apply to the high-SB inner regions of these galaxies.  However, these
\emph{are} the appropriate morphological K-corrections to be applied to
studies of high-redshift galaxies, where cosmological SB-dimming
similarly selects the high-SB regions as do our short \HST\ exposures. 

More statistics are needed to cover all late-types, and the major
dynamical stages of peculiars and mergers, as these are likely the
dominant population at high redshift.  We will pursue such objects
further through an \HST\ SNAP-shot program of late-type/irregular and
peculiar galaxies in Cycle 10.


\section{\bfseries\sc\large Summary and Conclusions}
\label{Conclusions}

In order to delineate the morphology of galaxies as observed in their
rest-frame mid-ultraviolet, we have carried out a systematic imaging
survey with \HST/WFPC2 of 37 nearby galaxies the \emph{mid-UV} filter
F300W (centered at 2930\AA), as well as in the $I$-band (filter F814W
centered at 8230\AA).  These mid-UV images will be useful to more
reliably classify the numerous faint galaxy images seen with \HST\ in
deep $I$-band surveys at high redshifts ($z\simeq1$--2).  Eleven
galaxies located in the \HST\ Continuous Viewing Zone were also imaged
in the F255W filter (centered at 2550\AA). 

Our sample is carefully selected for size and surface brightness, and
included galaxies of small enough radius and high enough predicted
mid-UV SB to be observable with \HST/WFPC2 in one orbit.  Together with
17 galaxies imaged in F300W available in the \HST\ Archive, our survey
covers a wide range of Hubble types and inclinations.  Complementary
data at other wavelengths are available for our galaxy sample: most
objects have ground-based $UBVR$ images, some have $IJHK$ images, and 15
have far-UV images from the \Astro/\UIT\ missions.  Mid-UV images
(2000--3200\AA) have been the missing ingredient thus far. 

In summary, our mid-UV imaging data base shows that in galaxies where
star-formation is sufficiently pronounced, it can dominate the
morphology from the mid-UV though the optical, resulting in very little
change in morphology from the UV to the red.  However, when stellar
populations older than about 1~Gyr produce most of the optical light, we
see changes in morphology between the optical --- where these stars tend
to produce relatively regular structures due to the effects of orbital
mixing within the galaxies --- and the mid-UV --- where younger stars
whose locations still reflect the distribution of their birthplaces.  An
additional complication is introduced by the presence of dust
obscuration.  Dust lanes or clouds that are nearly transparent in the
visible can be opaque in the mid-UV, thereby changing the apparent
morphology. 

As a function of intrinsic galaxy type, our first qualitative results
from this Cycle 9 project can be summarized as following:

\sn (1) Early-type galaxies (T=$-6$ to 0) show a significant change in
SB in going from the mid-UV to the red, reflecting their relative
paucity of a young stellar population.  Some early types appear rather
dim in the mid-UV due to significant central dust lanes, while others
show mostly a small blue nuclear feature in the mid-UV.  Three
ellipticals become nearly point sources in the mid-UV (\eg, dominated by
LINER's, Seyferts).  This is in part due to our sample selection, which
required the objects to have high (predicted) mid-UV SB.  But it leads
us to wonder to what extent the apparently strong cosmological evolution
of weak AGN in early-type galaxies is due to the morphological
K-correction plus SB-dimming, which causes the surrounding UV-faint
early-type host to no longer be visible at higher redshifts.  In
summary, high SB early-type galaxies in the optical show a variety of
morphologies in the mid-UV that {\it can} lead to a different
morphological classification, although not necessarily always as
later--type.  The often rather peculiar mid-UV morphology of early-type
galaxies is generally quite different than that of the real late-type
galaxies as seen in the mid-UV, as discussed below. 

\sn (2) Mid-type spiral galaxies (T=1--5) and star-forming galaxies can
appear as later morphological types in the mid-UV, as \Astro/\UIT\ has
shown primarily in the far-UV.  About half of the mid-type spirals in
the optical appear as later morphological types in the mid-UV, but not
all mid-type spiral galaxies do look dramatically different in the
mid-UV.  One mid-type spiral, NGC~6782 (\S~3.15), shows a spectacular
resonance ring full of hot stars in the F255W and F300W filters.  The
mid-UV images show a considerable range in the scale and SB of
individual star-forming regions.  A comparison of F300W to F814W images
yields good sensitivity to dust features.  Almost without exception, the
mid-type spirals in our sample have their small nuclear bulges bisected
by a dust-lane, which is often connected to the inner spiral arm
structure. 

\sn (3) The late-type, irregular, peculiar, and merging galaxies
(T=6--11) in our sample show diverse properties in the mid-UV.  The
majority of these galaxies show a F300W morphology that is similar to
that in F814W, but with important differences due to recognizable
dust-features absorbing the F300W light, and hot stars, star-clusters,
or star-formation ``ridges'' that are better visible in the mid-UV. 
Less than a third of galaxies classified as late-type in the optical
show significantly different structures in the mid-UV to result in a
different classification. 

In conclusion, our \HST\ mid-UV survey of nearby galaxies suggests that
it is more likely to misclassify true early--mid type galaxies in the
rest-frame mid-UV as later types, while known late-type galaxies are
less likely to be misclassified (as earlier types) in the mid-UV.  This
is because {\it truly} late-type galaxies are dominated by young and hot
stars in all filters from the mid-UV to the red, and so have to first
order the same morphology and a very small morphological K-correction in
general.  However, early-type galaxies (ellipticals and early-type
spirals) {\it can, although do not have to} look significantly different
when one goes from the rest-frame mid-UV to the optical-red part of the
spectrum. 

Classification of faint galaxies in the rest-frame mid-UV will thus
likely result in some fraction of early--mid type galaxies being
misclassified as later-types, probably a larger fraction than vice
versa.  But it is unlikely that this morphological K-correction can
explain all of the faint blue galaxy excess as misclassified
earlier-type galaxies.  And the morphological K-correction cannot
explain the slight excess of early--mid type galaxies at faint
magnitudes ($B\gsim 24$ mag) with respect to passively evolving models,
as found by Odewahn \etal (1996) and Cohen \etal (2002), since the main
misclassification error goes in the opposite direction.  Instead, our
mid-UV survey seems to support the conclusion of Cohen \etal (2002) that
the number of faint galaxies is larger than the non-evolving predictions
{\it for all galaxy types}, but more significantly so for the later
types.  Cohen \etal (2002) give a possible explanation of this finding
in terms of hierarchical formation models in a Lambda-dominated
Universe. 

To address these issues further quantitatively, we will in a sequel
paper use the Fourier decomposition method of Odewahn \etal (1995,
2002a), where we will quantify the morphological K-correction based on
each galaxy's mid-UV photometric parameters as a function of rest-frame
wavelength and observed type.  Our goal is to use the current \HST\
mid-UV sample to quantify the morphological K-correction, and apply it
to a large complete sample of faint \HST\ galaxies with know photometric
redshifts to get reliable and consistent rest-frame classifications. 
These will then be used to compute the redshift distribution as a
function of morphological type, which will help delineate the formation
and evolution of galaxies along the Hubble sequence (see Driver \etal
1998).


\acknowledgments

The US authors acknowledge support from NASA grants GO-8645.* and
AR-8765.*, and RSdJ acknowledges support from NASA through Hubble
Fellowship grant HF-01106.01-A, awarded by STScI, which is operated by
AURA for NASA under contract NAS 5-26555.  RAW, CAC and SCO acknowledge
support from NASA ADP grant NAG-6740.  CAC and VAT acknowledge support
from a NASA Space Grant Graduate Fellowship at ASU.  This research has
made use of the NASA/IPAC Extragalactic Database (NED), which is
operated by the Jet Propulsion Laboratory, California Institute of
Technology (Caltech) under contract with NASA, and has used NASA's
Astrophysics Data System Bibliographic Services.  The Second Palomar
Observatory Sky Survey (POSS-II) was made by Caltech with funds from the
NSF, NASA, the National Geographic Society, the Sloan Foundation, the
Samuel Oschin Foundation, and the Eastman Kodak Corporation.  The Oschin
Schmidt Telescope is operated by Caltech and Palomar Observatory.\\
We thank Tony Roman and the STScI staff for their excellent help in
getting these \HST\ observations scheduled.  We thank the staff of the
Vatican Advanced Technology Telescope, in particular Richard Boyle,
Chris Corbally, and Matt Nelson, for their excellent help in obtaining a
large fraction of the ground-based images presented in this paper.\\
We thank Seth Cohen for advice in the data reduction, Dr.\ Arthur Code
for a careful reading of the manuscript and for fruitful discussions,
and Prof.\ Peter Strittmatter and the University of Arizona for their
hospitality during a working visit.  We thank the referee, Dr.\ Rob
Kennicutt, for a careful and helpful review of the manuscript, and for
suggesting a new way to publish a paper like the current one.


\clearpage



\setcounter{figure}{0}

\begin{figure}[ht]
\epsscale{0.80}
\plotone{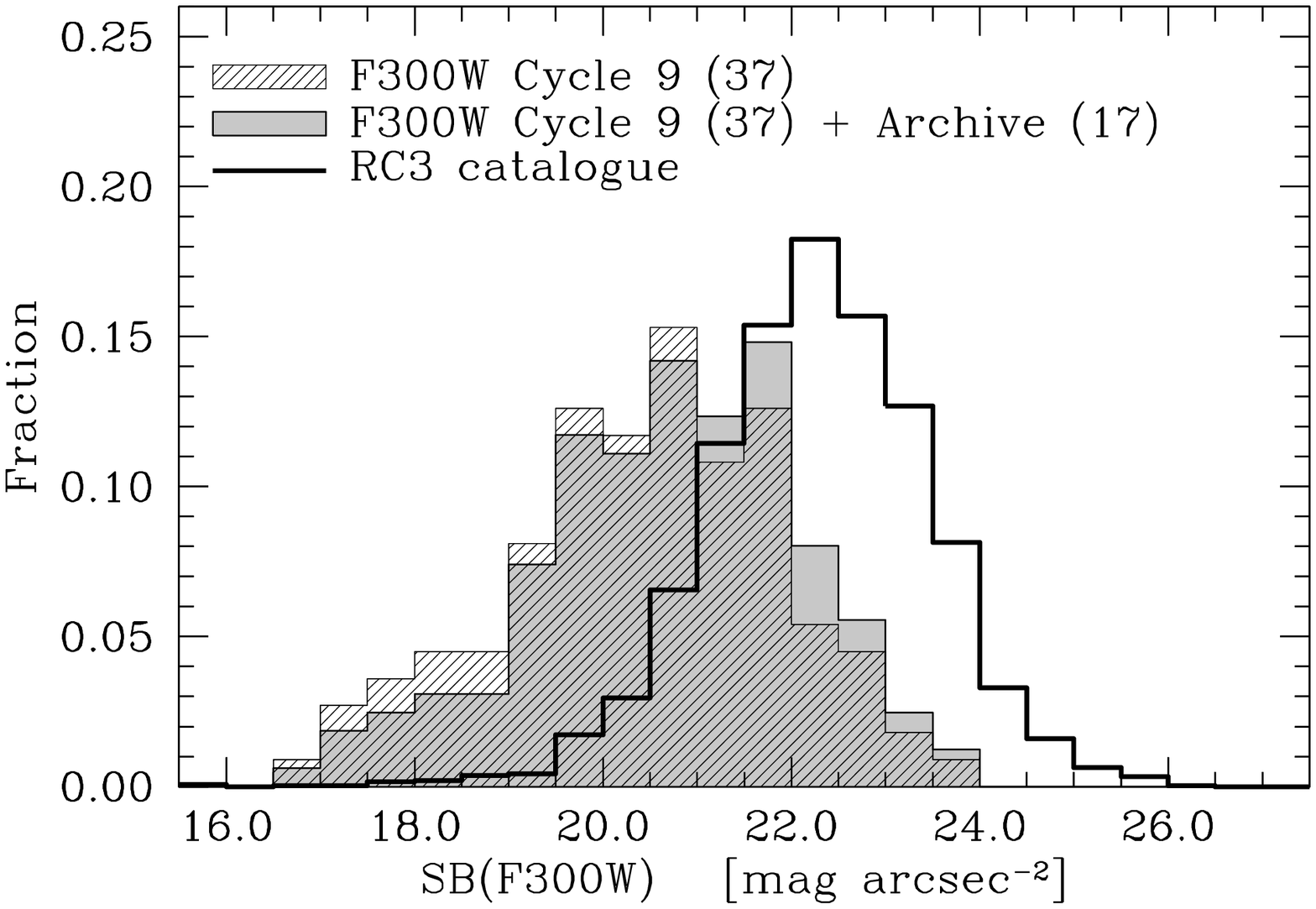}
\caption{The normalized distribution of the predicted average mid-UV
surface brightness (SB) out to \re\ (see \S~\ref{Strategy-PredictedSB})
for the 37 galaxies observed during \HST\ Cycle~9 (\emph{hashed
histogram}) and for the full galaxy sample (\emph{solid histogram}),
which includes 17 galaxies with mid-UV data taken prior to Cycle~9.  For
comparison, we also show the SB distribution for the 3009 galaxies in
the RC3 with measured $B_T$, ($U-B$) and \re\ (\emph{open histogram}). 
The galaxies were selected to have $\SB\ \lesssim 23$ \magarc, allowing
us to detect each object in F300W in no more than a single \HST\ orbit. 
For our purpose of comparing nearby and distant galaxies this imposed
SB-bias is justified, since the strong cosmological SB-dimming acts
similarly in hiding lower SB objects at high redshifts from deep
surveys.  \label{Figure-1} }
\end{figure}

\newpage

\begin{figure}[ht]
\epsscale{1.00}
\plottwo{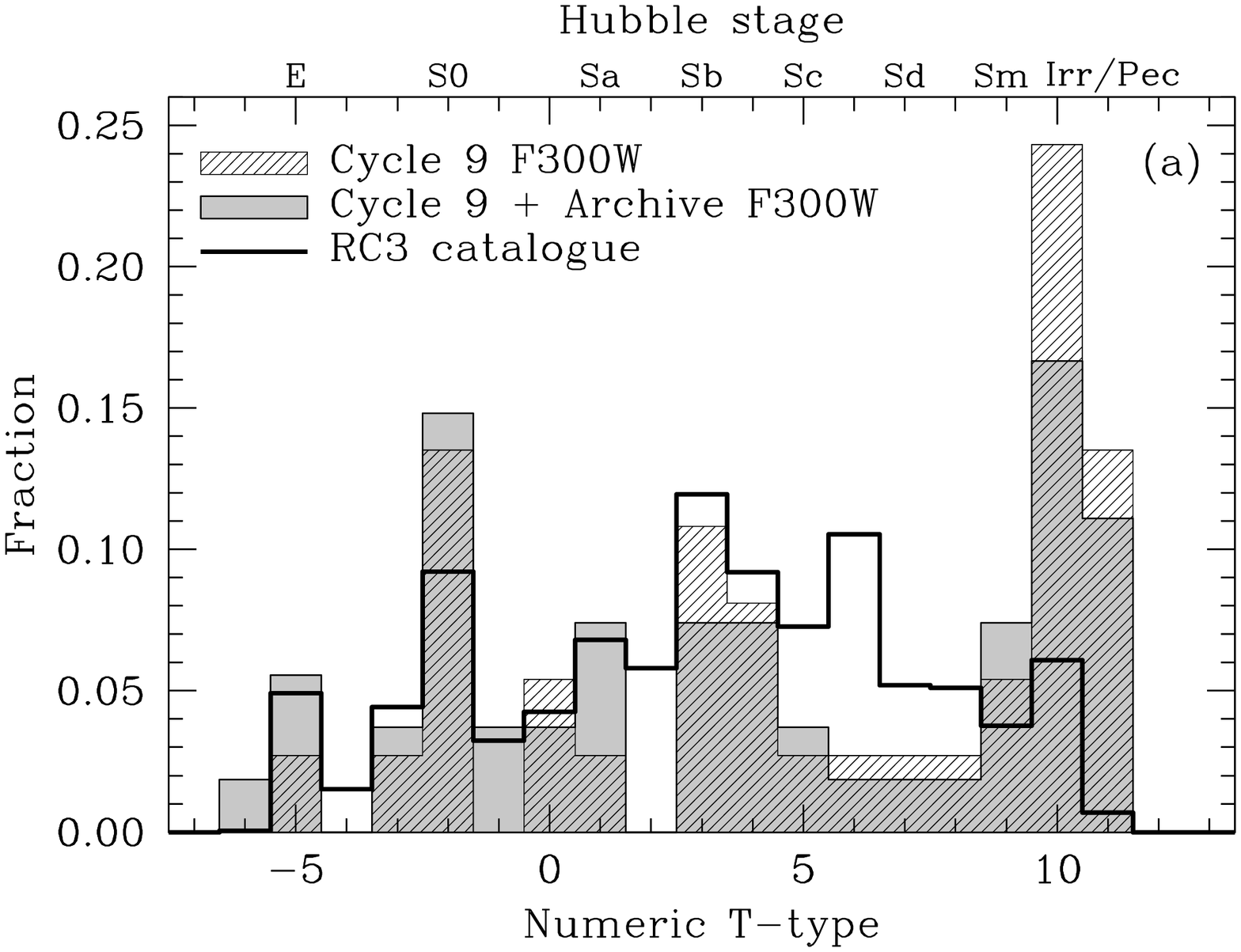}{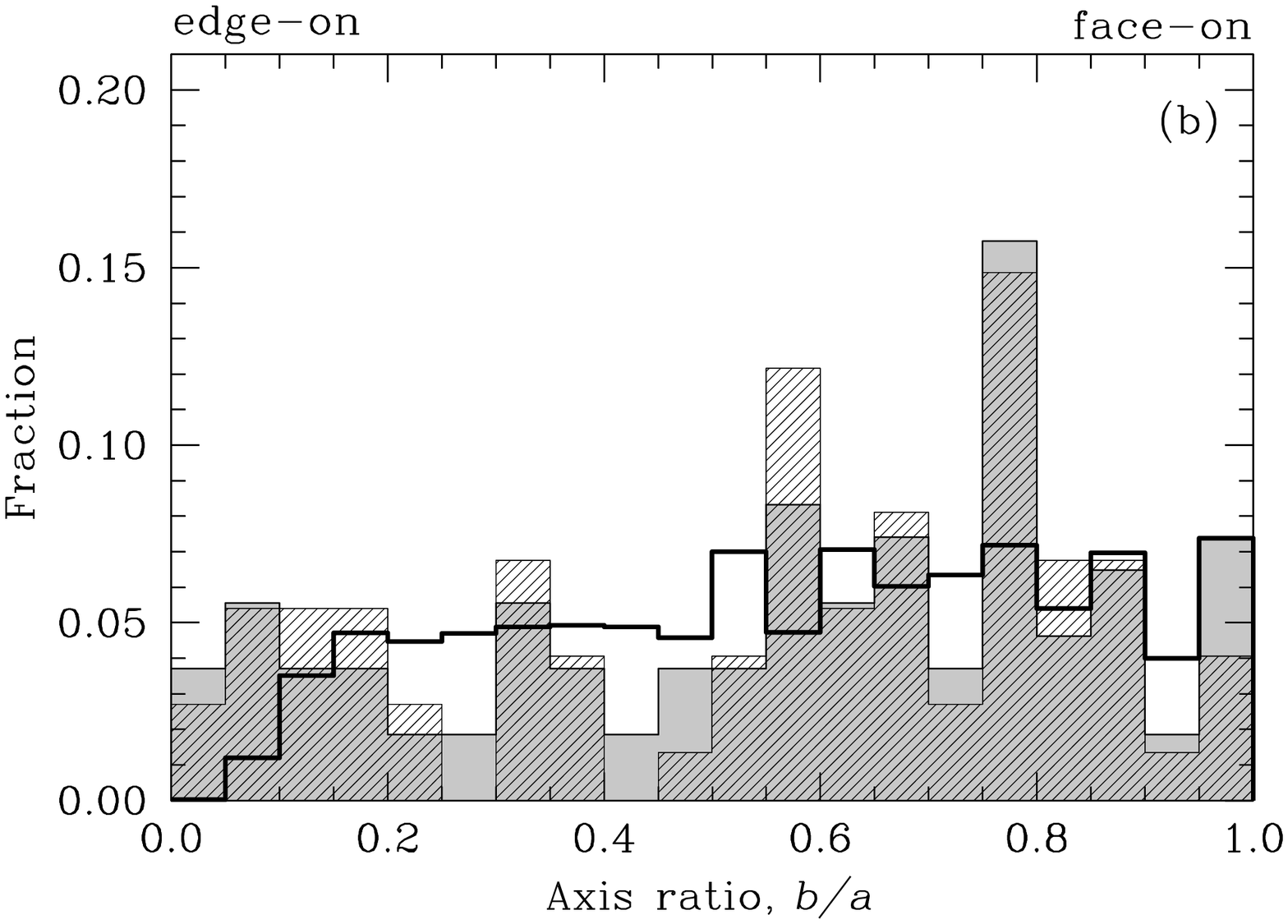}\\
\epsscale{2.25}
\plottwo{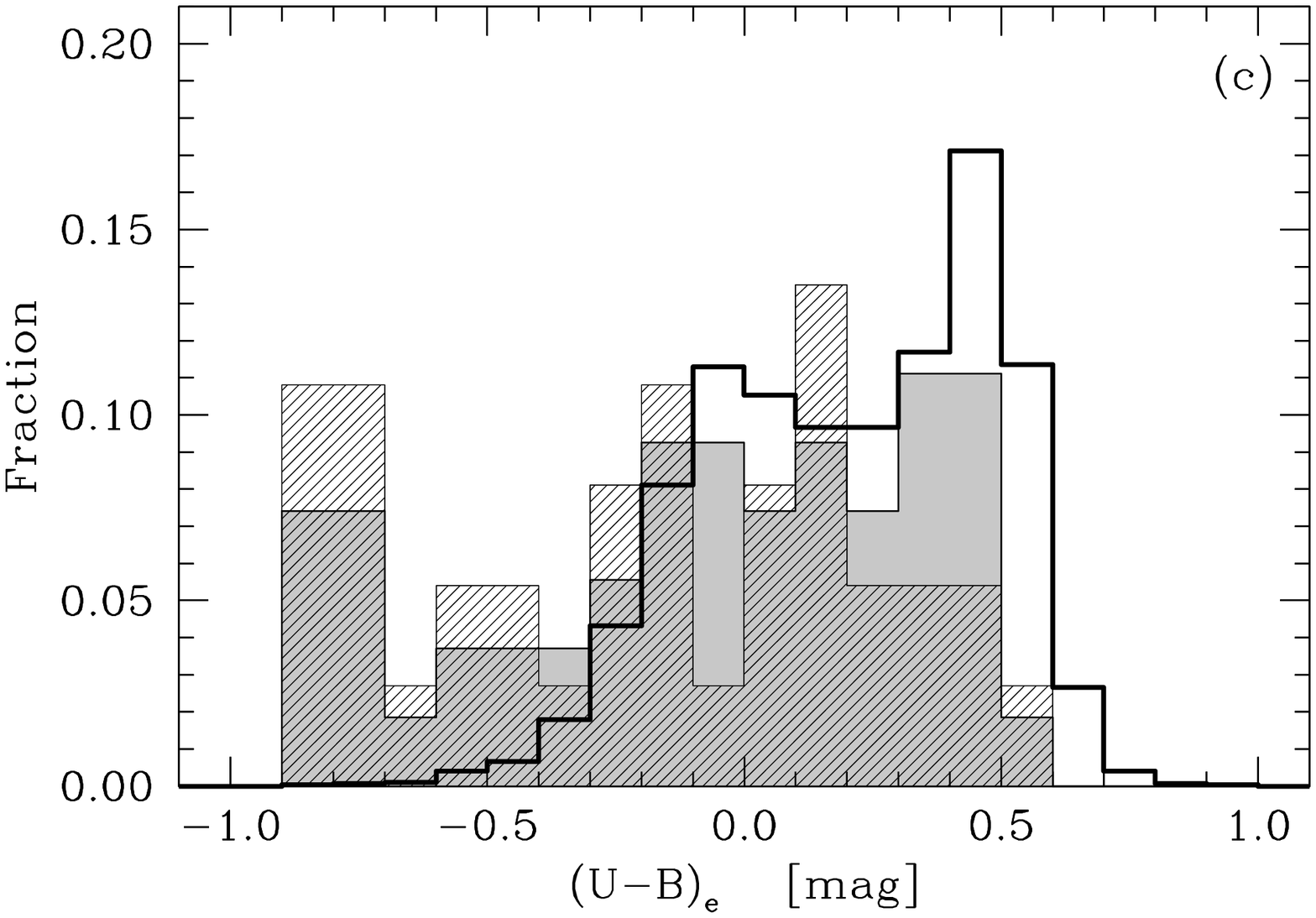}{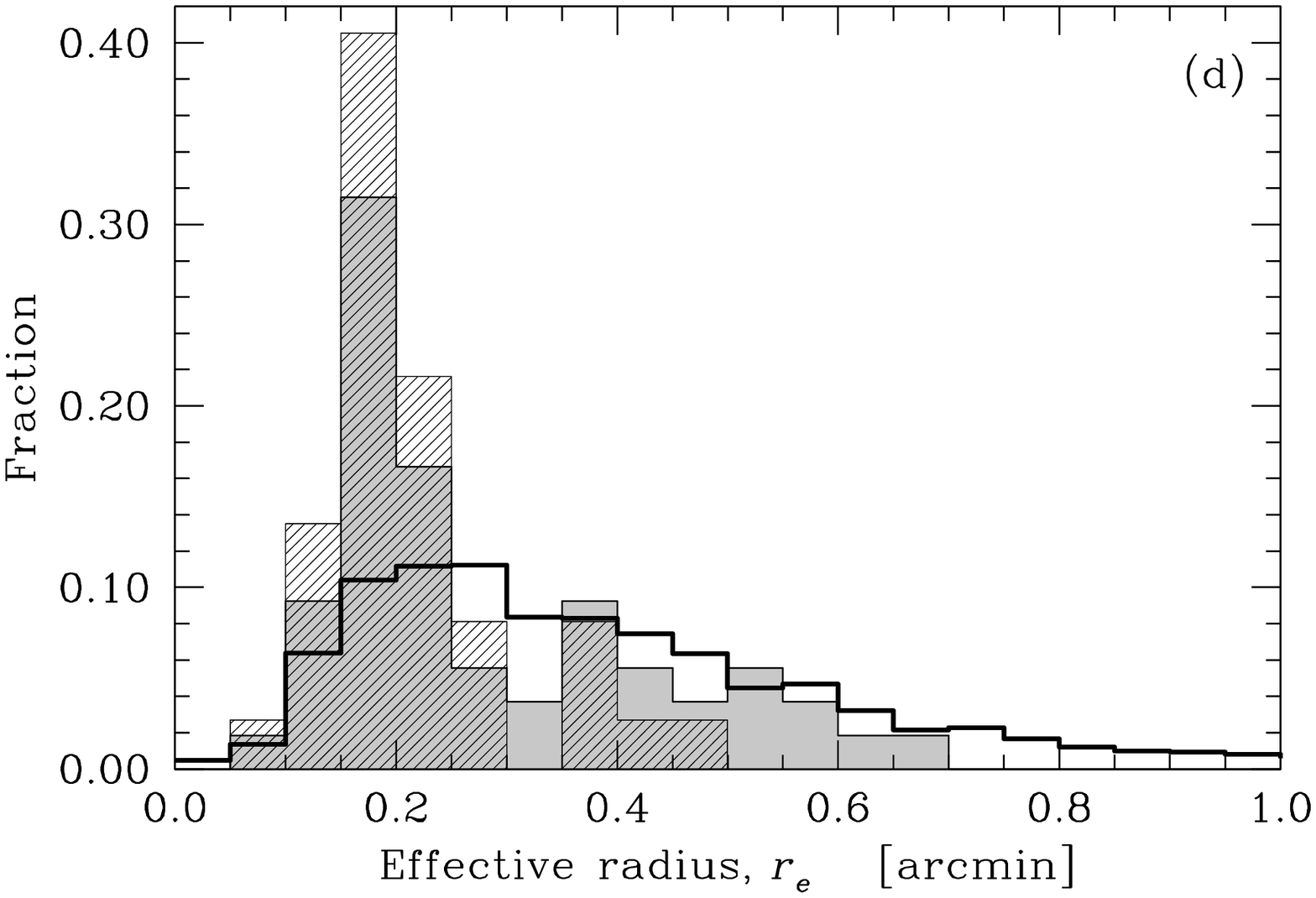}
\caption{Other properties of the selected galaxy sample.  As in
Fig.~1, we present normalized distributions for the Cycle~9 galaxies
(\emph{hashed}), the full sample (\emph{solid}), and the RC3
(\emph{open}) of (\emph{a}) morphological type; (\emph{b}) apparent axis
ratio, $b/a$; (\emph{c}) average ($U-B$) color out to \re; and
(\emph{d}) effective radius \re.  Taking deviations due to small number
statistics into account, our sample approximates the RC3 distribution,
except that for the purpose of comparison with high redshift objects, we
placed extra emphasis on the very latest types, bluest optical galaxy
colors, and smallest angular sizes (to fit the FOV of the WFPC2).  We
also somewhat over-represent highly inclined systems.
\label{Figure-2} }
\end{figure}

\newpage

\noindent
\framebox[\textwidth]{\rule[-4cm]{0pt}{8cm}
\large\bf See: fig3\emph{a}.jpg--fig3\emph{y}.jpg}

\figcaption{
Atlas of \HST\ WFPC2 observations of 37 nearby galaxies in
the mid-UV F255W (only for the 11 galaxies in \HST's CVZ) and F300W
filters, and in the optical $I$-band (F814W) filter.  The galaxies are
sorted in the atlas approximately according to their observed
morphological type following the discussion of the individual objects in
\S~\ref{NotesonIndividualObjects}.  For the convenience of the reader,
Table~2 cross-references the available galaxy data with the atlas page
numbering.  Each of the grey-scale renditions is displayed using a
logarithmic stretch and the image scale and orientation on the sky are
indicated.  For some galaxies we will present images at two different
image scales and stretches, in order to emphasize small-scale features
discussed in the text (see \S~\ref{NotesonIndividualObjects}).  We also
present images for some serendipitously observed galaxies.\\
--- \textsc{Mid-UV-faint elliptical galaxies.}\\ 
\textbf{3.01\ CGCG~097$-$094} (\emph{top}): WFPC2 F300W and F814W. 
Serendipitous observation of an early-type galaxy.  Also visible are an
edge-on spiral galaxy and a face-on disk galaxy (both anonymous);
\textbf{3.02\ NGC~1396} (\emph{bottom}): WFPC2 F300W and F814W.\\
\textsc{(cont'd)}
\textbf{3.03\ NGC~4478} WFPC2 F300W (\emph{left}) and F814W (\emph{right}).\\
\textsc{Early-type galaxies with significant AGN emission in the mid-UV.}\\
\textbf{3.04\ NGC~3516} (\emph{top}): WFPC2 F255W, F300W and F814W;
\textbf{3.05\ UGC~03426} (\emph{bottom}): WFPC2 F255W, F300W and F814W.\\
\textsc{(cont'd)}
\textbf{3.06\ UGC~08823} (\emph{top}): WFPC2 F300W and F814W.\\
\textsc{Early-type galaxies showing evidence for recent merging or
strong interaction.}\\
\textbf{3.07\ NGC~3921} (\emph{bottom}): WFPC2 F300W and F814W.\\
\textsc{(cont'd)}
\textbf{3.08\ UGC~05101} (\emph{top}): WFPC2 F255W, F300W and F814W;
\textbf{3.09\ UGC~08696} (\emph{bottom}): WFPC2 F255W, F300W and F814W.\\
\textsc{Early-type galaxies displaying spiral structure in the mid-UV.}\\
\textbf{3.10\ NGC~2551} WFPC2 F300W (\emph{left}) and F814W (\emph{right}).\\
\textsc{Grand-design spiral galaxies.}\\
\textbf{3.11a\ NGC~6753} (\emph{top}): WFPC2 F300W and F814W;
\textbf{3.11b\ NGC~6753} (\emph{bottom}): Nuclear portion of NGC~6753,
emphasizing the dust lane crossing the inner spiral arms discussed in
the text.\\
\textsc{(cont'd)}
\textbf{3.12a\ NGC~7769} (\emph{top}): WFPC2 F300W and F814W;
\textbf{3.12b\ NGC~7769} (\emph{bottom}): Central portion of NGC~7769,
emphasizing the dust lanes crossing the spiral arms and the dust lane
crossing its small bulge.\\
\textsc{(cont'd)}
\textbf{3.13a\ NGC~3310} (\emph{top}): WFPC2 F300W and F814W;
\textbf{3.13b\ NGC~3310} (\emph{bottom}): Nuclear portion of NGC~3310,
emphasizing the spiral arm and nucleus crossing dust lanes.\\
\textsc{(cont'd)}
\textbf{3.14\ NGC~7685} WFPC2 F300W (\emph{left}) and F814W (\emph{right}).\\
\textsc{Spiral galaxies with inner rings.}\\
\textbf{3.15a\ NGC~6782} (\emph{top}): WFPC2 F255W, F300W and F814W;
\textbf{3.15b\ NGC~6782} (\emph{bottom}): Enlargement of the central
portion of NGC~6782, emphasizing the spectacular ring of hot, young
stars.\\
\textsc{Edge-on spiral galaxies.}\\
\textbf{3.16\ UGC~10043} (\emph{top}): WFPC2 F300W and F814W;
\textbf{3.17\ UGC~06697} (\emph{bottom}): WFPC2 F300W and F814W.\\
\textsc{(cont'd)}
\textbf{3.18a\ IC~3949} (\emph{top}) WFPC2 F300W and F814W. Note the
four highly inclined systems visible in F814W that have their major axes
pointing toward IC~3949;
\textbf{3.18b\ IC~3949 (field)} (\emph{bottom}) Several serendipitously
observed early-type galaxies.\\
\textsc{(cont'd)}
\textbf{3.19\ ESO~033$-$G022} (\emph{top}): WFPC2 F300W and F814W;
\textbf{3.20\ IC~4393} (\emph{bottom}): WFPC2 F300W and F814W.\\
\textsc{Late-type spirals and smaller spirals.}\\
\textbf{3.21\ MCG+03$-$30$-$071} (\emph{top}): WFPC2 F300W and F814W.
Also visible is neighboring early-type galaxy MCG+03$-$30$-$067.
Comparison of the F300W and F814W images of these two galaxies gives
the best example of the general trends seen in the present work;
\textbf{3.22\ MCG+06$-$24$-$047} (\emph{bottom}): WFPC2 F300W and F814W.
The bright point source is a galactic (fore-ground) star.\\
\textsc{(cont'd)}
\textbf{3.23a\ UGC~05028} (\emph{top}): WFPC2 F300W and F814W;
\textbf{3.23b\ UGC~05029} (\emph{bottom}): WFPC2 F300W and F814W.\\
\textsc{(cont'd)}
\textbf{3.24\ NGC~386B} (\emph{top}): F300W and F814W;
\textbf{3.25\ ESO~418$-$G008} (\emph{bottom}): WFPC2 F300W and F814W.\\
\textsc{(cont'd)}
\textbf{3.26\ NGC~1510}: WFPC2 F300W (\emph{left}) and F814W (\emph{right}).\\
\textsc{Magellanic irregular galaxies.}\\
\textbf{3.27\ NGC~5253} (\emph{top}): WFPC2 F300W and F814W;
\textbf{3.28\ NGC~1140} (\emph{bottom}): WFPC2 F300W and F814W.\\
\textsc{(cont'd)}
\textbf{3.29\ UGC~05626}; WFPC2 F255W (\emph{left}), F300W
(\emph{middle}), and F814W (\emph{right}).\\
\textsc{(cont'd)}
\textbf{3.30\ UGC~09855} (\emph{top}): WFPC2 F255W, F300W and F814W;
\textbf{3.31\ NGC~6789} (\emph{bottom}): WFPC2 F255W, F300W and F814W.\\
\textsc{(cont'd)}
\textbf{3.32\ Mrk~66} (\emph{top}): WFPC2 F300W and F814W;
\textbf{3.33\ UGC~05189} (\emph{bottom}): WFPC2 F300W and F814W.\\
\textsc{Interacting, peculiar, and merging galaxies:}\\
\textbf{3.34a\ NGC~3690/IC~694} (\emph{top}): WFPC2 F255W, F300W and F814W;
\textbf{3.34b\ NGC~3690/IC~694} (\emph{bottom}): Enlargement of the central
portion of this merging galaxy pair, emphasizing the dusty patches and
star forming knots.\\
\textsc{(cont'd)}
\textbf{3.35\ UGC~06527} (\emph{top}): WFPC2 F300W and F814W;
\textbf{3.36\ UGC~08335} (\emph{bottom}): WFPC2 F300W and F814W.\\
\textsc{(cont'd)}
\textbf{3.37\ IC~2184} (\emph{top}): WFPC2 F255W, F300W, and F814W;
\textbf{3.38\ NGC~5278/79} (\emph{bottom}): WFPC2 F255W, F300W, and F814W.
\label{Figure-3}
}


\noindent
\framebox[\textwidth]{\rule[-4cm]{0pt}{8cm}
\large\bf See: fig4\emph{a}.jpg--fig4\emph{z}.jpg,
fig4\emph{aa}.jpg--fig4\emph{ll}.jpg}

\figcaption{ 
Pan-chromatic atlas of far- and mid-UV (\UIT), mid-UV (\HST/WFPC2), and
ground-based optical $UBVR$ and near-IR $JHK$ (ground or \HST/NICMOS)
observations of the 37 nearby galaxies in the present sample.  As in
Fig.~3, the galaxies are sorted and numbered according to morphology,
and each of the grey-scale renditions is displayed using a logarithmic
stretch.  Unlike Fig.~3, all images have been rotated to have North up
and East to the left, and are presented on a common plate scale.  The
printed version contains only a few example pages; for the full atlas we
refer the reader to the electronic version.\\
--- \textsc{Mid-UV-faint elliptical galaxies.}\\
\textbf{4.01\ CGCG~097$-$094} (\emph{top}): \UIT\ 150nm and 250nm;
(\emph{middle}): \HST/WFPC2 F300W, and ground-based $U$ and $B$;
(\emph{bottom}): ground-based $V$ and $R$, and \HST/WFPC2 F814W.\\
\textsc{(cont'd)}
\textbf{4.02\ NGC~1396} (\emph{top}): \UIT\ 150nm and 250nm, and
\HST/WFPC2 F300W; (\emph{bottom}): photographic $J$ (blue) and
IIIaF+RG610 (red), and \HST/WFPC2 F814W.\\
\textsc{(cont'd)}
\textbf{4.03\ NGC~4478} (\emph{top}): \UIT\ 150nm and 250nm;
(\emph{middle}): WFPC2 F300W, and ground-based $U$ and $B$;
(\emph{bottom}): ground-based $V$ and $R$, and \HST/WFPC2 F814W.\\
\textsc{Early-type galaxies with significant AGN emission in the mid-UV.}\\
\textbf{4.04\ NGC~3516} (\emph{top}): \UIT\ 150nm, and \HST/WFPC2 F255W;
(\emph{$2^{nd}$ row}): \HST/WFPC2 F300W, and ground-based $U$ and $B$;
(\emph{$3^{rd}$ row}): ground-based $V$ and $R$, and \HST/WFPC2 F814W;
(\emph{bottom}): \HST/NICMOS F160W ($H$).\\
\textsc{(cont'd)}
\textbf{4.05\ UGC~03426} (\emph{top}): \UIT\ 150nm and \HST/WFPC2 F255W;
(\emph{middle}): \HST/WFPC2 F300W, and ground-based $U$ and $B$;
(\emph{bottom}): ground-based $V$ and $R$, and \HST/WFPC2 F814W.\\
\textsc{(cont'd)}
\textbf{4.06\ UGC~08823} (\emph{top}): \UIT\ 150nm;
(\emph{middle}): \HST/WFPC2 F300W, and ground-based $U$ and $B$;
(\emph{bottom}): ground-based $V$ and $R$, and \HST/WFPC2 F814W.\\
\textsc{Early-type galaxies showing evidence for recent merging or
strong interaction.}\\
\textbf{4.07\ NGC~3921} (\emph{top}): \HST/WFPC2 F300W, and ground-based
$B$;
(\emph{middle}): ground-based $V$ and $R$, and \HST/WFPC2 F814W;
(\emph{bottom}): \HST/NICMOS F110W ($J$) and F160W ($H$), and
ground-based $K$.\\
\textsc{(cont'd)}
\textbf{4.08\ UGC~05101} (\emph{top}): \HST/WFPC2 F255W;
(\emph{$2^{nd}$ row}): \HST/WFPC2 F300W, and ground-based $U$ and $B$;
(\emph{$3^{rd}$ row}): ground-based $V$ and $R$, and \HST/WFPC2 F814W;
(\emph{bottom}): \HST/NICMOS F110W ($J$), F160W ($H$), and F220M ($K$).\\
\textsc{(cont'd)}
\textbf{4.09\ UGC~08696} (\emph{top}): \HST/WFPC2 F255W;
(\emph{$2^{nd}$ row}): \HST/WFPC2 F300W, and ground-based $U$ and
$IIIaJ+GG385$;
(\emph{$3^{rd}$ row}): ground-based $R$, and \HST/WFPC2 F814W;
(\emph{bottom}): \HST/NICMOS F110W ($J$), F160W ($H$), and F222M ($K$).\\
\textsc{Early-type galaxies displaying spiral structure in the mid-UV.}\\
\textbf{4.10\ NGC~2551} (\emph{top}): \UIT\ 150nm and 250nm;  
(\emph{middle}): \HST/WFPC2 F300W, and ground-based $U$ and $B$;
(\emph{bottom}): ground-based $V$ and $R$, and \HST/WFPC2 F814W.\\
\textsc{Grand-design spiral galaxies.}\\
\textbf{4.11\ NGC~6753} (\emph{top}): \HST/WFPC2 F300W, and ground-based
$U$ and $B$;
(\emph{middle}): ground-based $V$ and $R$, and \HST/WFPC2 F814W;
(\emph{bottom}): ground-based $J$, $H$, and $K$.\\
\textsc{(cont'd)}
\textbf{4.12\ NGC~7769} (\emph{top}): \HST/WFPC2 F300W, and ground-based
$U$ and $B$;
(\emph{middle}): ground-based $V$ and $R$, and \HST/WFPC2 F814W;
(\emph{bottom}): ground-based $I$ and $H$.\\
\textsc{(cont'd)}
\textbf{4.13\ NGC~3310} (\emph{top}): \UIT\ 150nm;
(\emph{$2^{nd}$ row}): \HST/WFPC2 F300W, and ground-based $U$ and $B$;
(\emph{$3^{rd}$ row}): ground-based $V$ and $R$, and \HST/WFPC2 F814W;
(\emph{bottom}): \HST/NICMOS F160W ($H$).\\
\textsc{(cont'd)}
\textbf{4.14\ NGC~7685} (\emph{top}): \HST/WFPC2 F300W, and ground-based
$U$ and $B$;
(\emph{middle}): ground-based $V$ and $R$, and \HST/WFPC2 F814W;
(\emph{bottom}): ground-based $I$ and $K$.\\
\textsc{Spiral galaxies with inner rings.}\\
\textbf{4.15\ NGC~6782} (\emph{top}): \HST/WFPC2 F255W;
(\emph{$2^{nd}$ row}): \HST/WFPC2 F300W, and ground-based $U$ and $B$;
(\emph{$3^{rd}$ row}): ground-based $V$ and $R$, and \HST/WFPC2 F814W;
(\emph{bottom}): ground-based $J$, $H$ and $K$.\\
\textsc{Edge-on spiral galaxies.}\\
\textbf{4.16\ UGC~10043} (\emph{top}): \HST/WFPC2 F300W, and
ground-based $U$ and $B$;
(\emph{bottom}): ground-based $V$ and $R$, and \HST/WFPC2 F814W.\\
\textsc{(cont'd)}
\textbf{4.17\ UGC~06697} (\emph{top}): \UIT\ 150nm and 250nm;
(\emph{middle}): \HST/WFPC2 F300W, and ground-based $U$ and $B$;
(\emph{bottom}): ground-based $V$ and $R$, and \HST/WFPC2 F814W.\\
\textsc{(cont'd)}
\textbf{4.18\ IC~3949} (\emph{top}): \UIT\ 150nm;
(\emph{middle}): \HST/WFPC2 F300W, and ground-based $U$ and $B$;
(\emph{bottom}): ground-based $V$ and $R$, and \HST/WFPC2 F814W.\\
\textsc{(cont'd)}
\textbf{4.19\ ESO~033$-$G022} (\emph{top}): \HST/WFPC2 F300W, and
ground-based $B$;
(\emph{bottom}): ground-based $V$ and $I$, and \HST/WFPC2 F814W.\\
\textsc{(cont'd)}
\textbf{4.20\ IC~4393} (\emph{top}): \HST/WFPC2 F300W, and
ground-based $U$ and $B$;
(\emph{middle}): ground-based $V$ and $I$, and \HST/WFPC2 F814W;
(\emph{bottom}): ground-based $I$ and $K$.\\
\textsc{Late-type spirals and smaller spiral galaxies.}\\
\textbf{4.21\ MCG+03$-$30$-$071} (\emph{top}): \UIT\ 150nm and 250nm;
(\emph{middle}): \HST/WFPC2 F300W, and ground-based $U$ and $B$;
(\emph{bottom}): ground-based $V$ and $I$, and \HST/WFPC2 F814W.\\
\label{Figure-4}
}

\begin{minipage}[ht]{\textwidth}\footnotesize
\textsc{(cont'd)}
\textbf{4.22\ MCG+06$-$24$-$047} (\emph{top}): \UIT\ 150nm;
(\emph{middle}): \HST/WFPC2 F300W, and ground-based $U$ and $B$;
(\emph{bottom}): ground-based $V$ and $I$, and \HST/WFPC2 F814W.\\
\textsc{(cont'd)}
\textbf{4.23\ UGC~05028/29} (\emph{top}): \HST/WFPC2 F300W, and 
ground-based $IIIaJ+GG385$;
(\emph{bottom}): ground-based $IIIaF+RG610$, and \HST/WFPC2 F814W.\\
\textsc{(cont'd)}
\textbf{4.24\ NGC~3860B} (\emph{top}): \UIT\ 150nm and 250nm;
(\emph{middle}): \HST/WFPC2 F300W, and ground-based $U$ and $B$;  
(\emph{bottom}): ground-based $V$ and $R$, and \HST/WFPC2 F814W.\\
\textsc{(cont'd)}
\textbf{4.25\ ESO~418$-$G008} (\emph{top}): \HST/WFPC2 F300W, and 
ground-based $U$ and $B$;
(\emph{bottom}): ground-based $V$ and $R$, and \HST/WFPC2 F814W.\\
\textsc{(cont'd)}
\textbf{4.26\ NGC~1510} (\emph{top}): \UIT\ 150nm;
(\emph{middle}): \HST/WFPC2 F300W, and ground-based $B$ (from Lauberts
\& Valentijn 1989, as obtained through NED);
(\emph{bottom}): ground-based $IIIaF+RG610$, and \HST/WFPC2 F814W.\\
\textsc{Magellanic irregular galaxies.}\\
\textbf{4.27\ NGC~5253} (\emph{top}): \UIT\ 150nm, and \HST/WFPC2 F255W;
(\emph{$2^{nd}$  row}): \HST/WFPC2 F300W, and ground-based $U$ and $B$;
(\emph{$3^{rd}$  row}): ground-based $V$ and $R$, and \HST/WFPC2 F814W;
(\emph{bottom}): ground-based $J$, $H$, and $K$.\\
\textsc{(cont'd)}
\textbf{4.28\ NGC~1140} (\emph{top}): \HST/WFPC2 F300W, and ground-based
$U$ and $B$;
(\emph{bottom}): ground-based $V$ and $IIIaF+RG610$, and \HST/WFPC2
F814W.\\
\textsc{(cont'd)}
\textbf{4.29\ UGC~05626} (\emph{top}): \HST/WFPC2 F255W;
(\emph{middle}): \HST/WFPC2 F300W, and ground-based $U$ and $B$;
(\emph{bottom}): ground-based $V$ and $R$, and \HST/WFPC2 F814W.\\
\textsc{(cont'd)}
\textbf{4.30\ UGC~09855} (\emph{top}): \HST/WFPC2 F255W;
(\emph{middle}): \HST/WFPC2 F300W, and ground-based $U$ and $B$;
(\emph{bottom}): ground-based $V$ and $R$, and \HST/WFPC2 F814W.\\
\textsc{(cont'd)}
\textbf{4.31\ NGC~6789} (\emph{top}): \HST/WFPC2 F255W;
(\emph{middle}): \HST/WFPC2 F300W, and ground-based $U$ and $B$;
(\emph{bottom}): ground-based $V$ and $R$, and \HST/WFPC2 F814W.\\
\textsc{(cont'd)}
\textbf{4.32\ Mrk~66} (\emph{top}): \UIT\ 150nm;
(\emph{middle}): \HST/WFPC2 F300W, and ground-based $U$ and $B$;
(\emph{bottom}): ground-based $V$ and $R$, and \HST/WFPC2 F814W.\\
\textsc{(cont'd)}
\textbf{4.33\ UGC~05189} (\emph{top}): \HST/WFPC2 F300W, and
ground-based $U$ and $B$;
(\emph{bottom}): ground-based $V$ and $R$, and \HST/WFPC2 F814W.\\
\textsc{Interacting, peculiar, and merging galaxies.}\\
\textbf{4.34\ NGC~3690/IC~694} (\emph{top}): \HST/WFPC2 F255W;
(\emph{$2^{nd}$ row}): \HST/WFPC2 F300W, and ground-based $U$ and $B$;
(\emph{$3^{rd}$ row}): ground-based $V$ and $R$, and \HST/WFPC2 F814W;
(\emph{bottom}): \HST/NICMOS F110M ($J$) and F160W ($H$), and
ground-based $K$.\\
\textsc{(cont'd)}
\textbf{4.35\ UGC~06527} (\emph{top}): \HST/WFPC2 F300W, and 
ground-based $U$ and $B$;
(\emph{middle}): ground-based $V$ and $R$, and \HST/WFPC2 F814W;
(\emph{bottom}): \HST/NICMOS F160W ($H$).\\
\textsc{(cont'd)}
\textbf{4.36\ UGC~08335} (\emph{top}): \HST/WFPC2 F300W, and
ground-based $U$ and $IIIaJ+GG385$;
(\emph{bottom}): ground-based $R$, and \HST/WFPC2 F814W.\\
\textsc{(cont'd)}
\textbf{4.37\ IC~2184} (\emph{top}): \HST/WFPC2 F255W;
(\emph{middle}): \HST/WFPC2 F300W, and ground-based $U$ and $B$;
(\emph{bottom}): ground-based $V$ and $R$, and \HST/WFPC2 F814W.\\
\textsc{(cont'd)}
\textbf{4.38\ NGC~5278/79} (\emph{top}): \HST/WFPC2 F255W;
(\emph{middle}): \HST/WFPC2 F300W, and ground-based $U$ and
$IIIaJ+GG385$;
(\emph{bottom}): ground-based $R$, and \HST/WFPC2 F814W.
\end{minipage}


\setlength{\tabcolsep}{4pt}
\begin{deluxetable}{lcccccccccccrcll}
\tabletypesize{\scriptsize}
\tablecolumns{16}
\tablewidth{6.8in}
\tablenum{1}
\tablecaption{Nearby galaxies imaged with WFPC2 in the mid-UV in
              Cycle~9\,$^a$} 
\tablehead{
\colhead{Galaxy}       & \multicolumn{2}{c}{RA\ \ \ (J2000)\ \ \ DEC}&
\colhead{$b/a$}        & \colhead{$r_e$}      & \colhead{$r_{25}$}   &
\colhead{SB$_e^{U300}$}& \colhead{SB$_{r25}^{U300}$}& \colhead{$U-B$}&
\colhead{$B-V$}        & \colhead{$B$}        & \colhead{$U300$}     &
\colhead{T}            & \colhead{$z$}        & \colhead{Scale}      &
\colhead{kpc}          \\
\colhead{}             & \colhead{}           & \colhead{}           &
\colhead{}             & \colhead{(\arcmin)}  & \colhead{(\arcmin)}  &
\multicolumn{2}{c}{(mag/arcsec$^2$)}          & \colhead{(mag)}      &
\colhead{(mag)}        & \colhead{(mag)}      & \colhead{(mag)}      &
\colhead{}             & \colhead{}           & \colhead{(pc/\arcsec)} &
\colhead{/FOV}         \\
\colhead{(1)}          & \colhead{(2)}        & \colhead{(3)}        &
\colhead{(4)}          & \colhead{(5)}        & \colhead{(6)}        &
\colhead{(7)}          & \colhead{(8)}        & \colhead{(9)}        &
\colhead{(10)}         & \colhead{(11)}       & \colhead{(12)}       &
\colhead{(13)}         & \colhead{(14)}       & \colhead{(15)}       &
\colhead{(16)}         }
\startdata
\multicolumn{16}{l}{\sc --- Cycle 9 WFPC2 mid-UV Sample:}\\
NGC 1140                 & 02:54:33.43 & $-$10:01:42.4 & 0.55 & 0.144 & 0.830 & 18.29 & 21.34 & $-$0.43 & 0.35    & 12.84 & 12.26 &   10.0 & 0.0049 & 106.~ & ~7.95 \\
ESO 418$-$G008           & 03:31:30.58 & $-$30:12:46.6 & 0.66 & 0.185 & 0.587 & 20.61 & 22.37 & $-$0.23 & 0.41    & 13.92 & 13.84 &   ~8.0 & 0.0038 & ~82.4 & ~6.18 \\
NGC 1396                 & 03:38:06.63 & $-$35:26:24.5 & 0.10 & 0.172 & 0.500 & 21.19 & 22.75 &  ~~0.53 & \nodata & 14.80 & 16.62 & $-$3.0 & 0.0026 & ~56.4 & ~4.23 \\
NGC 1510                 & 04:03:32.55 & $-$43:24:03.0 & 0.55 & 0.151 & 0.659 & 19.62 & 22.07 & $-$0.19 & 0.45    & 13.47 & 13.49 & $-$2.3 & 0.0028 & ~60.7 & ~4.55 \\
ESO 033$-$G022           & 05:31:41.58 & $-$73:45:04.2 & 0.06 & 0.248 & 1.021 & 19.92 & 22.24 & $-$0.30 & \nodata & 15.37 & 15.12 &   ~7.0 & 0.0138 & 299.~ & 22.4~ \\
UGC 03426 (Mrk 3)        & 06:15:36.33 & $+$71:02:13.8 & 0.87 & 0.085 & 0.910 & 20.36 & 24.76 &  ~~0.18 & 1.06    & 14.03 & 14.98 & $-$2.0 & 0.0138 & 299.~ & 22.4~ \\
\enddata
\vspace*{-4mm}
\footnotesize
\tablecomments{(a) Unless stated otherwise, all table entries are taken
or derived from de RC3 catalogue (de Vaucouleurs \etal 1991).\\
\null\hspace*{0.9em}{\sc Notes on table columns} --- (1) galaxy name;
(2) and (3) J2000 (RA, DEC) coordinates of the galaxy centers as
measured on the Digital Sky Survey (DSS).  In a few cases of complex
galaxies, galaxy mergers, or galaxies in small groups, the listed
coordinates refer to a ``center of gravity position'', allowing optimal
centering inside the WFPC2 FOV; (4) ratio of the minor and major axis
isophotal diameter at the SB$(B)=25.0$ \magarc\ isophote; (5) $r_e$
denotes the half-light radius; (6) $r_{25}$ denotes the radius at the
SB$(B)=25.0$ \magarc\ isophote; (7) predicted average SB in F300W out to
$r_e$ (see section 2.2); (8) predicted average SB in F300W out to
$r_{25}$ (see \S2.2); (9) measured Johnson $(U-B)$ aperture color; (10)
measured $(B-V)$ aperture color; (11) total $B$-band magnitude; (12)
predicted total F300W magnitude (see \S2.2); (13) revised galaxy T-type
(see also Fig.~2a); (14) Galactocentric galaxy redshift, as derived from
the RC3; (15) number of parsec per arcsec at the distance of the galaxy,
assuming a Hubble constant of 67 \kmsMpc; (16) physical size in kpc
corresponding to a single WFPC2 CCD of 75\arcsec\ [Col.\ 16 = Col.\ 
$15\times 0.075$). }

\end{deluxetable}
\normalsize

\setlength{\tabcolsep}{8pt}
\begin{deluxetable}{lllclcclll}
\tabletypesize{\scriptsize}
\tablecolumns{10}
\tablewidth{6.8in}
\tablenum{2}
\tablecaption{Summary of \HST, \UIT\ and ground-based images of our
              Cycle~9 sample}
\tablehead{
\colhead{Object}       & \colhead{Telescope}   & \colhead{Filter}     &
\colhead{Exposure time}& \colhead{Airmass}     & \colhead{Orient}     &
\colhead{Zeropoint}    & \colhead{Plate scale} & \colhead{FWHM}       &
\colhead{Fig.nr.}      \\
\colhead{}             & \colhead{}            & \colhead{}           &
\colhead{(sec)}        & \colhead{}            & \colhead{(deg)}      &
\colhead{(mag)}        & \colhead{(\arcsec/pix)}& \colhead{(\arcsec)} &
\colhead{}             \\
\colhead{(1)}          & \colhead{(2)}         & \colhead{(3)}        &
\colhead{(4)}          & \colhead{(5)}         & \colhead{(6)}        &
\colhead{(7)}          & \colhead{(8)}         & \colhead{(9)}        &
\colhead{(10)}         }
\startdata
NGC 1140 (Mrk 1063)   & HST      & F300W  & 2$\times$~900.0 & 0     &  71.933 & 19.433 & 0.0996 & 0.04 & 3/4.28 \\
                      & HST      & F814W  & 2$\times$~100.0 & 0     &  71.933 & 21.659 & 0.0996 & 0.08 & 3/4.28 \\
                      & VATT     & U      & 1$\times$~300.0 & 1.364 &   0.0   &   ---  & 0.3746 & 2.15 & 4.28   \\
                      & VATT     & B      & 1$\times$~180.0 & 1.360 &   0.0   &   ---  & 0.3746 & 2.59 & 4.28   \\
                      & VATT     & V      & 1$\times$~180.0 & 1.358 &   0.0   &   ---  & 0.3746 & 2.49 & 4.28   \\
                      & POSS-II  & IIIaF  & 1$\times$3000.0 & ---   &   0.0   &   ---  & 1.00   & 3.86 & 4.28   \\[2pt]
ESO 418-G008          & HST      & F300W  & 2$\times$~900.0 & 0     & 151.410 & 19.433 & 0.0996 & 0.04 & 3/4.25 \\
                      & HST      & F814W  & 2$\times$~130.0 & 0     & 151.410 & 21.659 & 0.0996 & 0.08 & 3/4.25 \\
\enddata
\end{deluxetable}
\normalsize

\setlength{\tabcolsep}{6pt}
\begin{deluxetable}{llrrlrlll}
\tabletypesize{\scriptsize}
\tablecolumns{9}
\tablewidth{6.8in}
\tablenum{3}
\tablecaption{Other imaging data in the \HST\ Archive on our Cycle~9
	galaxy sample.}
\tablehead{
\colhead{Object}         & \colhead{Filter}         &
\colhead{$\lambda_{c}$}  & \colhead{$\Delta\lambda$}&
\colhead{Exposure time}  & \colhead{Orient}         &
\colhead{Proposal ID}    & \colhead{PI}             &
\colhead{Note\hfil\null} \\
\colhead{}               & \colhead{}               &
\colhead{(\AA)}          & \colhead{(\AA)}          &
\colhead{(sec)}          & \colhead{(deg)}          &
\colhead{}               & \colhead{}               &
\colhead{}               \\
\colhead{(1)}            & \colhead{(2)}            &
\colhead{(3)}            & \colhead{(4)}            &
\colhead{(5)}            & \colhead{(6)}            &
\colhead{(7)}            & \colhead{(8)}            &
\colhead{(9)\hfil\null}  }
\startdata
\textsc{--- WFPC2 Images:}\\
ESO 418-G008   & F606W & 5957 & 1508 & 2$\times$80    & 227.216 & SNAP~5446 & G. Illingworth    & center in PC1 \\
               & F814W & 7940 & 1531 & 2$\times$600   &  16.701 & SNAP~8599 & T. Boeker         & PC1 \\
NGC 1510       & F170W & 1730 &  500 & 2$\times$400   & 303.570 & GO~6716   & T. Stecher        &  \\
               & F336W & 3344 &  381 & 2$\times$260   & 303.570 & ""        & ""                &  \\
               & F555W & 5407 & 1236 & 1$\times$200   & 303.570 & ""        & ""                &  \\
               & F814W & 7940 & 1531 & 1$\times$200   & 303.570 & ""        & ""                &  \\
UGC 03426      & F606W & 5957 & 1508 & 1$\times$500   &  73.315 & SNAP~5479 & M. Malkan         & PC1 \\
\enddata
\end{deluxetable}
\normalsize

\end{document}